\documentclass[preprint2,trackchanges]{aastex6}
\shorttitle{Magnetic Properties of Flaring Solar ARs}
\shortauthors{Toriumi et al.}
\begin{document}


\title{Magnetic Properties of Solar Active Regions that Govern Large Solar Flares and Eruptions}

%% Use \author, \affil, plus the \and command to format author and affiliation 
%% information.  If done correctly the peer review system will be able to
%% automatically put the author and affiliation information from the manuscript
%% and save the corresponding author the trouble of entering it by hand.
%%
%% The \affil should be used to document primary affiliations and the
%% \altaffil should be used for secondary affiliations, titles, or email.

%% Authors with the same affiliation can be grouped in a single
\author{Shin Toriumi$^{1,}$\altaffilmark{2}, Carolus J. Schrijver$^{3}$, Louise K. Harra$^{4}$,
  Hugh Hudson$^{5}$, Kaori Nagashima$^{6}$}
\affil{$^{1}$National Astronomical Observatory of Japan, 2-21-1 Osawa, Mitaka, Tokyo 181-8588, Japan\\
$^{3}$Lockheed Martin Advanced Technology Center, 3251 Hanover Street, Palo Alto, CA 94304, USA\\
$^{4}$UCL-Mullard Space Science Laboratory, Holmbury St Mary, Dorking, Surrey, RH5 6NT, U.K.\\
$^{5}$SUPA School of Physics and Astronomy, University of Glasgow, U.K.\\
$^{6}$Max-Planck-Institut f\"{u}r Sonnensystemforschung, Justus-von-Liebig-Weg 3, 37077 G\"{o}ttingen, Germany}
\altaffiltext{2}{shin.toriumi@nao.ac.jp}

%% Use the \and command so offset the last author.
% \and

% \author{Jeff Lewandowski\altaffilmark{5}}
% \affil{IOP Publishing, Washington, DC 20005}


% \altaffiltext{1}{AAS Journals Data Scientist}
% \altaffiltext{2}{greg.schwarz@aas.org}
% \altaffiltext{3}{AAS Journals Associate Editor-in-Chief}
% \altaffiltext{4}{AAS Director of Publishing}
% \altaffiltext{5}{IOP Senior Publisher for the AAS Journals}

\begin{abstract}
Solar flares and coronal mass ejections (CMEs),
especially the larger ones,
emanate from active regions (ARs).
With the aim to understand
the magnetic properties
that govern such flares and eruptions,
we systematically survey
all flare events
with GOES levels
of $\ge$ M$5.0$
within 45$^{\circ}$ from disk center
between May 2010 and April 2016.
These criteria lead to
a total of 51 flares from 29 ARs,
for which
we analyze the observational data
obtained by the {\it Solar Dynamics Observatory}.
More than 80\%
of the 29 ARs
are found to
exhibit $\delta$-sunspots
and at least three ARs
violate Hale's polarity rule.
The flare durations
are approximately proportional
to the distance between the two flare ribbons,
to the total magnetic flux inside the ribbons,
and to the ribbon area.
From our study,
one of the parameters
that clearly determine
whether a given flare event is
CME-eruptive or not
is the ribbon area
normalized by the sunspot area,
which may indicate that
the structural relationship
between the flaring region
and the entire AR
controls
CME productivity.
AR characterization show that even
X-class events do not require
$\delta$-sunspots or
strong-field,
high-gradient polarity inversion lines.
An investigation of historical observational data
suggests the possibility that
the largest solar ARs,
with magnetic flux of $2\times 10^{23}\ {\rm Mx}$,
might be able to produce
``superflares''
with energies
of order of $10^{34}\ {\rm erg}$.
The proportionality
between the flare durations and
magnetic energies
is consistent with stellar flare observations,
suggesting a common physical background
for solar and stellar flares.
\end{abstract}




\section{Introduction}\label{sec:introduction}

Solar flares
and coronal mass ejections (CMEs)
are the most catastrophic
energy-conversion phenomena
in the present solar system.
It is now widely accepted that
flares are
associated with
magnetic reconnection,
a physical process
that rearranges
the magnetic
configuration
and converts magnetic energy
into kinetic energy
and thermal energy,
and with
non-thermal particle acceleration
\citep{pri02,shi11}.
Observations show that
similar flaring phenomena
are found
in a wide variety
of stars
\citep{ben10}.

Since their discovery
by \citet{car59} and \citet{hod59},
solar flares,
especially the larger ones,
are known to appear
in and around active regions (ARs)
including sunspots.
Observationally,
the complex ARs
called $\delta$-sunspots,
in which umbrae of positive and negative polarities
share a common penumbra,
tend to produce
larger flare eruptions
\citep{kue60,zir87,sam00}.
In the $\delta$-spots,
the neighboring polarities
are likely to possess
a strong-field, high-gradient,
highly-sheared polarity inversion line (PIL),
which indicates
the existence
of intense currents
that can store free magnetic energy
above in the corona
\citep{sch07}.

As the flare evolves,
the two ribbons
extend around the sheared PIL,
which is observed
in H$\alpha$ and other chromospheric lines
\citep[e.g.,][]{dod49a,bru64,asa04}.
In the standard model for eruptive flares, the CSHKP model
\citep{car64,stu66,hir74,kop76},
the flare ribbons are caused by
magnetic reconnection
through the precipitation of high-energy electrons
and the effect of thermal conduction,
and thus they indicate
the footpoints of
newly reconnected field lines
(post-flare loops).

CMEs are often associated with flares,
particularly with the more energetic ones
\citep[e.g.,][]{web87,and03,yas05,hud10}.
The studies by,
e.g., \citet{wan07}, \citet{che11},
\citet{kah15}, \citet{tha15}, and \citet{sun15}
suggest a trend
that CME-eruptive flares occur
at larger distances
from the AR centers
and have larger ``decay index'' values
\citep{kli06}
than the non-eruptive ones
(also referred to as confined flares
or failed eruptions).
Although a general picture begins
to emerge from these studies,
each of them is based on
a small sample of selected flares
from different regions
(some along with non-flaring counterparts)
or on a small set of events
from one target region.
In a recent analysis,
\citet{har16}
used the set of X-class flares
from the current solar cycle
to show that
there is no obvious difference
in the flare duration
between the eruptive and non-eruptive events
and that the non-eruptive ones
tend to have larger spot area.
In this study,
we expand the sample
used by \citet{har16}
by lowering the threshold to mid-M-class flares
and by expanding the time interval
(see Section \ref{sec:selection}).
This also enables us
to test the findings
from the studies
referenced earlier
in this paragraph
using a much larger sample
in which the only selection bias
is that they do not occur
too far from disk center,
so that their magnetic patterns
are well observed.

One of the ultimate goals
of this paper is
to find the physical parameters
that dictate
the peak magnitudes and time scales
of the {\it GOES}
({\it Geostationary Orbiting Environmental Satellites})
soft X-ray (SXR) flux
(GOES parameters)
and that determine
whether a given flare
becomes CME-eruptive or not,
and in case of eruption,
the speed of the CME
(CME parameters).
For this purpose,
we carry out a systematic survey
of observational data of flaring ARs
obtained by the {\it Solar Dynamics Observatory}
({\it SDO}; \citealt{pes12}).
We especially evaluate
the morphological properties
of such flaring ARs
for characterizing
the magnetic environment
of flare eruptions
(AR parameters)
as well as those of the flare ribbons
as indicators of magnetic fields
in flare reconnection
(flare parameters),
and investigate the statistical relationships
among the GOES parameters, CME parameters,
AR parameters, and flare parameters.

Another aim is to elucidate
the formation process of flaring ARs.
Since such large-scale ARs are
created by flux emergence,
the transportation
of dynamo-generated magnetic flux
in the deeper interior
to the solar surface
\citep{par55},
it is of necessity
to conduct numerical simulations
and model the flux emergence
\citep{fan09,che14},
which is, however,
beyond the scope of this paper.
Instead,
we here
characterize
the flaring ARs
by focusing on the positional relationship
between
the flare-related ribbons (closely associated with the PIL)
and the entire AR,
and examine the statistical tendencies.

To understand
the diversity of flare events,
such as extreme solar events
in history,
simply surveying
the {\it SDO} data set
may not be sufficient.
One possible way to fulfill this desire
is to learn from the past.
Therefore,
as an example,
we also introduce
a great flare event
that occurred in 1946
(SOL1946-07-25)
and explore
the possibilities
of historical data analysis
in combination
with our statistical results.

The rest of the paper
is organized as follows.
In Sections \ref{sec:analysis}
and \ref{sec:results},
we describe the data analysis
and show the statistical results,
respectively.
Then,
Section \ref{sec:classification}
is dedicated to
characterizing
the flaring ARs
and their statistical trends,
while in Section \ref{sec:historical},
we show the analysis
on our historical flare event.
We discuss the results
in Section \ref{sec:discussion},
and finally
in Section \ref{sec:conclusion},
we conclude the paper.

\section{Data Analysis}\label{sec:analysis}

\subsection{Event Selection and Parameters}\label{sec:selection}

For the purpose of this study,
we analyzed all flare events
with GOES soft X-ray (SXR) magnitudes
greater than or equal to M5.0
within $45^{\circ}$ from the disk center
in the period from May 2010 to April 2016,
namely,
six years from the beginning
to the declining phase
of Solar Cycle 24.
These constraints
led us to
a total of 51 flares from 
29 regions with distinct
NOAA AR numbers.
Table \ref{tab:flares1} summarizes
the target 51 flare events.
For each event,
we obtained these GOES parameters
from the GOES SXR (1 -- 8\ {\AA} channel) light curve:
\begin{itemize}
 \item full width at half maximum (FWHM) of the light curve,
   $\tau_{\rm FWHM}$,
 \item $e$-folding decay time,
   $\tau_{\rm decay}$, and
 \item peak flux, $F_{\rm SXR}$.
\end{itemize}
We give the method of estimation of
the two time scales
in Section \ref{sec:sample}.

In order to identify
the physical parameters
that may characterize
the flare events,
we used the observational data
taken by the Helioseismic and Magnetic Imager
(HMI; \citealt{sche12,scho12})
and the Atmospheric Imaging Assembly
(AIA; \citealt{lem12})
aboard {\it SDO}.
For each event,
we made projection-corrected tracked data cubes
of the intensitygram,
line-of-sight (LOS) magnetogram,
and 1600\ {\AA} images,
using the {\tt mtrack} module
\citep{bog11}.
From the intensitygram and 1600\ {\AA} data,
center-to-limb variations were subtracted
based on the method
introduced in \citet{tor14b}.
The data cubes have
a cadence of 180 s for magnetogram and intensitygram
and 120 s for 1600\ {\AA} images,
both having pixel size
$1\farcs 0$.
The sequence of each data cube
is 2 or 3 hours
starting from before
the GOES start time
(typically 20 minutes before),
while the area of the field of view (FOV)
is $512\arcsec\times 256\arcsec$
or $1024\arcsec\times 512\arcsec$
centering the target AR.

From the first frames
of the intensitygram and magnetogram
of each flare event,
which is well before
the flare peak
(typically 40 minutes before),
we selected
the following AR parameters:
\begin{itemize}
 \item sunspot area, $S_{\rm spot}$,
   which is the total area
   of umbrae and penumbrae of an AR,
   defined in this study
   as the de-projected area of the pixels
   with intensity
   less than 85\%
   of the mean quiet-Sun intensity,   
 \item total unsigned flux,
   $|\Phi|_{\rm AR}=\int_{S_{\rm FOV}}|B|\,ds$,
   where
   $S_{\rm FOV}\,(=512\arcsec\times 256\arcsec$
   or $1024\arcsec\times 512\arcsec)$ is the FOV
   area\footnote{
     Depending on the target AR,
     we used a FOV
     of $512\arcsec\times 1024\arcsec$
     or $1024\arcsec\times 2048\arcsec$
     to cover it.
     However, in some cases,
     the rectangular FOV
     contains
     the neighboring flux concentrations
     that may not be related to
     the target region.
     We masked such flux concentrations
     to obtain a better AR area
     (see Figure \ref{fig:all}).},
   $B$ is the LOS magnetic flux density in each pixel,
   and $s\,(=1\arcsec\times 1\arcsec)$ is the pixel area,
   and
 \item normalized field strength,
   which is the total AR flux normalized
   by the spot area,
   $\overline{|B|}_{\rm AR}=|\Phi|_{\rm AR}/S_{\rm spot}$.
\end{itemize}
Here,
the spot areas were measured
in millionths of the solar hemisphere (MSH),
which is equivalent
to $3.0\times 10^{6}\ {\rm km}^{2}$.

For each event,
we extracted
the flare ribbons
from 1600\ {\AA} images
by defining them
as the pixels
with intensity
at any time during the flare
(until the last frame)
equal to or larger than
40$\sigma$
(standard deviation)
above the mean
of the quiet-Sun values,
and made binary ribbon maps.
After removing the saturated frames,
we stacked the binary ribbon maps over time
and made a binary ribbon composite.
By plotting the ribbon composite
over the magnetogram (first frame),
we divide the composite
into two parts,
the ribbon in the positive polarity
and that in the negative polarity.
Then,
the flare parameters were defined as:
\begin{itemize}
  \item ribbon area,
    $S_{\rm ribbon}$,
    the total area
    of the ribbon composite,
  \item ribbon distance,
    $d_{\rm ribbon}$,
    the separation
    between the two
    area-weighted (i.e., geometrical)
    centroids of the ribbons
    in the positive and negative polarities,
  \item total unsigned flux
    inside the ribbon,
    $|\Phi|_{\rm ribbon}=\int_{S_{\rm ribbon}}|B|\,ds$, and
  \item normalized field strength of the ribbon,
    $\overline{|B|}_{\rm ribbon}=|\Phi|_{\rm ribbon}/S_{\rm ribbon}$.
\end{itemize}

In addition,
we determined
whether each flare event
was CME-eruptive or not
(i.e., confined)
by reference to the
CME catalog\footnote{\url{http://cdaw.gsfc.nasa.gov/CME_list/}} of
the {\it Solar and Heliospheric Observatory} ({\it SOHO})/Large Angle Spectroscopic Coronagraph (LASCO).
For eruptive events,
\begin{itemize}
 \item CME speed, $V_{\rm CME}$,
 the linear speed obtained
 by fitting a straight line
 to the height-time measurements,
\end{itemize}
was also listed
from the catalog
as a CME parameter.
Note that
because
the current analysis
is based only on the flare events
within $45^{\circ}$ from the disk center,
there is a potential
to miss some CMEs:
\citet{yas05} suggests that
roughly one in six CMEs
are missed
from the on-disk M-class events
(see Table 3 of their paper).

Furthermore,
we followed the long-term evolution
of each AR
by making HMI data cubes
that covers
the AR's whole disk passage.

\subsection{Sample Event}\label{sec:sample}

Figure \ref{fig:sample} shows
an example of
the analyzed data sets:
the X3.1-class event in NOAA AR 12192.
From the first frames
of the HMI intensitygram and magnetogram
(panels (a) and (b)),
we measured
the spot area $S_{\rm spot}$,
total flux $|\Phi|_{\rm AR}$,
and normalized field strength
$\overline{|B|}_{\rm AR}$.
By temporally stacking the flare ribbons
extracted from the AIA 1600\ {\AA} images
(panel (c)),
we made a ribbon composite
(panel (d)),
which provides
our measure of
the ribbon area $S_{\rm ribbon}$.
Then,
by overlaying the ribbon composite
on the magnetogram
(panel (e)),
we measured
the ribbon distance
$d_{\rm ribbon}$,
the total flux $|\Phi|_{\rm ribbon}$,
and the mean field strength
$\overline{|B|}_{\rm ribbon}$.
Panel (f) shows the GOES SXR (1 -- 8 {\AA}) light curve.
For measuring the FWHM time $\tau_{\rm FWHM}$,
the background level,
which is the flux
at the GOES start time,
is first subtracted from the light curve.
The $e$-folding decay time $\tau_{\rm decay}$
is calculated using the flux $F_{\rm SXR}(t)$
and its time derivative $dF_{\rm SXR}(t)/dt$
at the GOES end time
as $\tau_{\rm decay}=-F_{\rm SXR}(t)/(dF_{\rm SXR}(t)/dt)$.

Since the flare ribbons expand
as the flare evolves
(Section \ref{sec:introduction}),
we need to take into account the effect of this expansion,
especially the timing when the evolution slows.
Figure \ref{fig:evolution} compares
the GOES light curve,
the evolution of the ribbon composite area
$S_{\rm ribbon}(t)$
(ribbon composite
made from the AIA 1600 {\AA} data sets
until each moment $t$),
and the evolution of the ribbon distance
$d_{\rm ribbon}(t)$
(ribbon distance measured from $S_{\rm ribbon}(t)$).
The final values of these parameters
are used in the analysis
as $S_{\rm ribbon}$ and $d_{\rm ribbon}$.
In the middle panel,
we measure
the {\it actual} ribbon area
at each moment $t$
and overplot it as 
$S^{\ast}_{\rm ribbon}(t)$.

Although the most impulsive period
is not seen
due to saturation
in the 1600\ {\AA} images,
$S^{\ast}_{\rm ribbon}(t)$
reaches its maximum
in the rising phase
before the GOES peak time.
This is reasonable
because the ribbons indicate
the heating
of chromospheric plasma
via thermal conduction
and high-energy electrons
driven by the reconnection,
while the SXR loops
are formed
following
chromospheric evaporation
\citep[see, e.g.,][]{shi11}.
This may be
one manifestation
of the so-called Neupert effect
\citep{neu68}.
As a result,
images showing
the evolutions of
$S_{\rm ribbon}(t)$
and thus $d_{\rm ribbon}(t)$
become saturated
around the GOES peak.

In this study,
the end times
of the {\it SDO} data sets
are well after the GOES peak times,
and thus we can consider
that the ribbon composite
of each flare event
sufficiently reflects
the expansion of
bright ribbons.

\section{Statistical Results}\label{sec:results}

\subsection{Properties of ARs and Flare Events}\label{sec:properties}

Table \ref{tab:flares1} shows
the Mount Wilson sunspot classification.
Here,
23 out of the 29 ARs
($79$\%)
show a $\delta$-structure
at least for one flare occurrence.
However,
although AR 11158 was reported as non-$\delta$,
this region actually shows
a $\delta$-configuration
when it produces the flares.
Therefore,
the actual fraction
increases to
83\%
for the $\ge$M5 events
under study.
This result is in line with previous results
that the
$\delta$-spots have
higher flare productivity
\citep[e.g.,][]{sam00}.

There are
three ARs
(10\%)
that violate Hale's polarity rule
for at least one flare
(ARs 11429, 11719, and 12158).
If we also count
AR 12242,
which shows anti-Hale structure
until about one day before the flare eruption,
this fraction becomes
14\%.
Although this number
is much larger
than the typically reported value
of 3 -- 5\%
for all ARs
\citep{ric48,wan89,khl09},
the small sample number
does not allow
any firm conclusion about this.

The analyzed 51 flares are composed of
20 X- and 31 M-class events,
ranging from M5.0 to X5.4.
They include several major flares
from well-studied ARs.
Among others,
NOAA AR 11158 produced
the first X-class (X2.2) flare
in Solar Cycle 24
\citep[e.g.,][]{sch11},
11429 produced
the largest (X5.4) flare so far
in this cycle
\citep[e.g.,][]{wan14},
12017 produced
the ``best-observed'' X1.0-class flare
\citep[e.g.,][]{kle15},
and 12192, the largest sunspot group
so far in the cycle,
produced many (6 X- and 24 M-class)
but CME-poor events
(e.g., \citealt{sun15}:
4 X- and 2 M-class events are listed
in Table \ref{tab:flares1}).
Almost all the events
in this table
occurred at PILs
within the AR's magnetic structure itself.
However,
there are two exceptional cases:
events \#29 (X1.2) and \#34 (M5.1) occurred
at the PIL
between two neighboring ARs.

\subsection{Parameters that Dictate GOES Light Curves}\label{sec:goes}

In this study,
from the {\it SDO} data set
of each flare event,
we measured various parameters:
GOES parameters
(durations $\tau_{\rm FWHM}$ and $\tau_{\rm decay}$
and GOES flux $F_{\rm SXR}$),
AR parameters
(spot area $S_{\rm spot}$,
total flux $|\Phi|_{\rm AR}$,
and field strength $\overline{|B|}_{\rm AR}$),
and flare parameters
(ribbon area $S_{\rm ribbon}$,
distance $d_{\rm ribbon}$,
total flux $|\Phi|_{\rm ribbon}$,
and field strength $\overline{|B|}_{\rm ribbon}$).
The values for all
events
and their maximum, minimum, median, and standard deviation values
are shown
in Table \ref{tab:flares2}
of Appendix \ref{app:list}.
But here
we list
the ranges and medians
of these parameters:
$\tau_{\rm FWHM}=154$ -- 4790 s
(median: 1198 s),
$\tau_{\rm decay}=32$ -- 1986 s
(433 s),
$F_{\rm SXR}=(0.5$ -- $5.4)\times 10^{-4}\ {\rm W\ m^{-2}}$
($0.87\times 10^{-4}\ {\rm W\ m}^{-2}$),
$S_{\rm spot}=126$ -- 2877 MSH
(781 MSH),
$|\Phi|_{\rm AR}=(1.1$ -- $16.6)\times 10^{22}\ {\rm Mx}$
($3.8\times 10^{22}\ {\rm Mx}$),
$\overline{|B|}_{\rm AR}=568$ -- 810 G
(685 G),
$S_{\rm ribbon}=102$ -- 1639 MSH
(431 MSH)
$d_{\rm ribbon}=4.1$ -- 105.1 Mm
(26.9 Mm),
$|\Phi|_{\rm ribbon}=(0.9$ -- $16.1)\times 10^{21}\ {\rm Mx}$
($4.4\times 10^{21}\ {\rm Mx}$),
and
$\overline{|B|}_{\rm ribbon}=125$ -- 590 G
(308 G).

In order to find the physical parameters
that dictate the GOES light curves,
we made scatter plots
of the measured data,
namely,
the scatter plots of
$y=\{\tau_{\rm FWHM}, \tau_{\rm decay}, F_{\rm SXR}\}$
versus $x=\{S_{\rm spot}, |\Phi|_{\rm AR}, \overline{|B|}_{\rm AR},
S_{\rm ribbon}, d_{\rm ribbon}, |\Phi|_{\rm ribbon}, \overline{|B|}_{\rm ribbon}\}$.
For $x$,
we also used
the ratio of the two areas,
$S_{\rm ribbon}/S_{\rm spot}$,
ranging 9.0 -- 328\%
(median: 56\%),
and that of total fluxes,
$|\Phi|_{\rm ribbon}/|\Phi|_{\rm AR}=1.6$ -- 43\%
(11\%).
Then,
for obtaining the empirical relationship,
we evaluated for each diagram
the power-law index $\alpha$
by fitting the data with a power-law function,
$\log{y}=\alpha\log{x}+{\rm const.}$,
or $y\propto x^{\alpha}$.
We also measured
the correlation coefficient,
$CC(\log{x},\log{y})$,
to estimate the degree of dispersion
of each plot.
Note that we assumed
errors for $y$-coordinate only.

As a result,
we obtained 27 scatter plots
and thus 27 empirical relations,
whose power-law indices $\alpha$
and correlation coefficients $CC$
are summarized
in Table \ref{tab:power}.
Figure \ref{fig:param} displays
the six least-scattered plots
(strongest correlations with $|CC|\ge 0.6$:
highlighted with bold face
in Table \ref{tab:power}).

The best correlations are obtained
from the scatter plots
of the FWHM duration of the flares,
$\tau_{\rm FWHM}$
(Figures \ref{fig:param}(a--c)).
They are of the flare parameters:
the ribbon distance,
$d_{\rm ribbon}$,
\begin{eqnarray}
  \log{\tau_{\rm FWHM}}=(0.96\pm0.09)\log{d_{\rm ribbon}}+(1.67\pm 0.13),
  \nonumber \\
  \label{eq:tfwhm_dribbon}
\end{eqnarray}
the ribbon total flux,
$|\Phi|_{\rm ribbon}$,
\begin{eqnarray}
  \log{\tau_{\rm FWHM}}=(1.04\pm0.12)\log{|\Phi|_{\rm ribbon}}+(-19.4\pm 2.51),
  \nonumber \\
  \label{eq:tfwhm_pribbon}
\end{eqnarray}
and the ribbon area,
$S_{\rm ribbon}$,
\begin{eqnarray}
  \log{\tau_{\rm FWHM}}=(1.10\pm0.15)\log{S_{\rm ribbon}}+(0.08\pm 0.40),
  \nonumber \\
  \label{eq:tfwhm_sribbon}
\end{eqnarray}
with correlation coefficients
of $CC=0.83$, 0.79, and 0.72,
respectively.
Interestingly,
all the above equations show
power-law indices
of approximately unity,
$\alpha\sim 1$.
The other parameters
show a more scattered distribution:
except for the AR field strength,
$\overline{|B|}_{\rm AR}$,
they have positive relations
(see Table \ref{tab:power}).

The other three best relations
are of the $e$-folding decay time,
$\tau_{\rm decay}$
(Figures \ref{fig:param}(d--f)),
and they are
of the same flare parameters:
the ribbon distance, $d_{\rm ribbon}$,
\begin{eqnarray}
  \log{\tau_{\rm decay}}=(0.88\pm0.12)\log{d_{\rm ribbon}}+(1.35\pm 0.18),
  \nonumber \\
  \label{eq:tdecay_dribbon}
\end{eqnarray}
the ribbon total flux, $|\Phi|_{\rm ribbon}$,
\begin{eqnarray}
  \log{\tau_{\rm decay}}=(0.96\pm0.15)\log{|\Phi|_{\rm ribbon}}+(-18.2\pm 3.21),
  \nonumber \\
  \label{eq:tdecay_pribbon}
\end{eqnarray}
and the ribbon area, $S_{\rm ribbon}$,
\begin{eqnarray}
  \log{\tau_{\rm decay}}=(1.05\pm0.18)\log{S_{\rm ribbon}}+(-0.21\pm 0.47),
  \nonumber \\
  \label{eq:tdecay_sribbon}
\end{eqnarray}
with $CC=0.71$, 0.68, and 0.64,
respectively.
It is natural that
$\tau_{\rm decay}$ also shows
strong correlations
with the above three parameters,
because $\tau_{\rm FWHM}$ and $\tau_{\rm decay}$
are highly correlated with each other
($CC=0.87$).
Although
the distributions
for $\tau_{\rm decay}$
are a bit more scattered
and thus the correlations are
slightly weaker
than those of $\tau_{\rm FWHM}$,
the power-law indices still show
$\alpha\sim 1$.
The other parameters
also show similar trends to
those of $\tau_{\rm FWHM}$
with similar power-law indices.
However, again,
the correlations are on average
weaker than
those of $\tau_{\rm FWHM}$.

On the other hand,
no diagrams
of the GOES peak flux,
$F_{\rm SXR}$,
have higher ($|CC|\ge 0.6$) correlations.
The maximum correlation coefficient here
is just $CC=0.37$
of the ribbon total flux,
$|\Phi|_{\rm ribbon}$.
They show generally positive correlations,
but $\overline{|B|}_{\rm AR}$
and $S_{\rm ribbon}/S_{\rm spot}$ show
negative relations.

In this data set,
we only have a range
of one order of magnitude
for the GOES peak flux,
$F_{\rm SXR}=(0.5$ -- $5.4)\times 10^{-4}\ {\rm W\ m}^{-2}$,
while the GOES durations
span more than one order,
$\tau_{\rm FWHM}=154$ -- $4790\ {\rm s}$
and
$\tau_{\rm decay}=32$ -- $1986\ {\rm s}$.
This narrow range of $F_{\rm SXR}$
may be one of the factors
that cause the weaker correlations.

\subsection{Parameters that Determine CME Properties}\label{sec:cme}

In our data set
of 51 $\ge$M5-class events,
there are 32 CME eruptive
and 19 non-eruptive events.
In this section,
we search
the parameters that determine
CME rich/poor
and their speed.

Figure \ref{fig:cmeyn} displays
the histograms
for CME eruptive and non-eruptive events.
From the top row,
one may see that
there is no significant difference
in distributions
of durations and magnitudes
between the eruptive and non-eruptive cases.
The averages of the log values
(indicated by vertical dashed lines)
for the eruptive and non-eruptive are
$\tau_{\rm FWHM}=1068$ and 826 s
(difference $=26$\%\footnote{Hereinafter,
we use relative difference
$|a-b|/(|a+b|/2)$
to show the quantitative difference
between $a$ and $b$.}),
$\tau_{\rm decay}=386$ and 314 s
(20\%),
and $F_{\rm SXR}=1.1\times 10^{-4}$
and $0.83\times 10^{-4}\ {\rm W\ m}^{-2}$
(23\%),
respectively.
Thus,
at least for the $\ge$M5-class events,
the longer-duration or larger-magnitude flares
are not necessarily CME-eruptive.

\begin{figure}
  \begin{center}
    \includegraphics{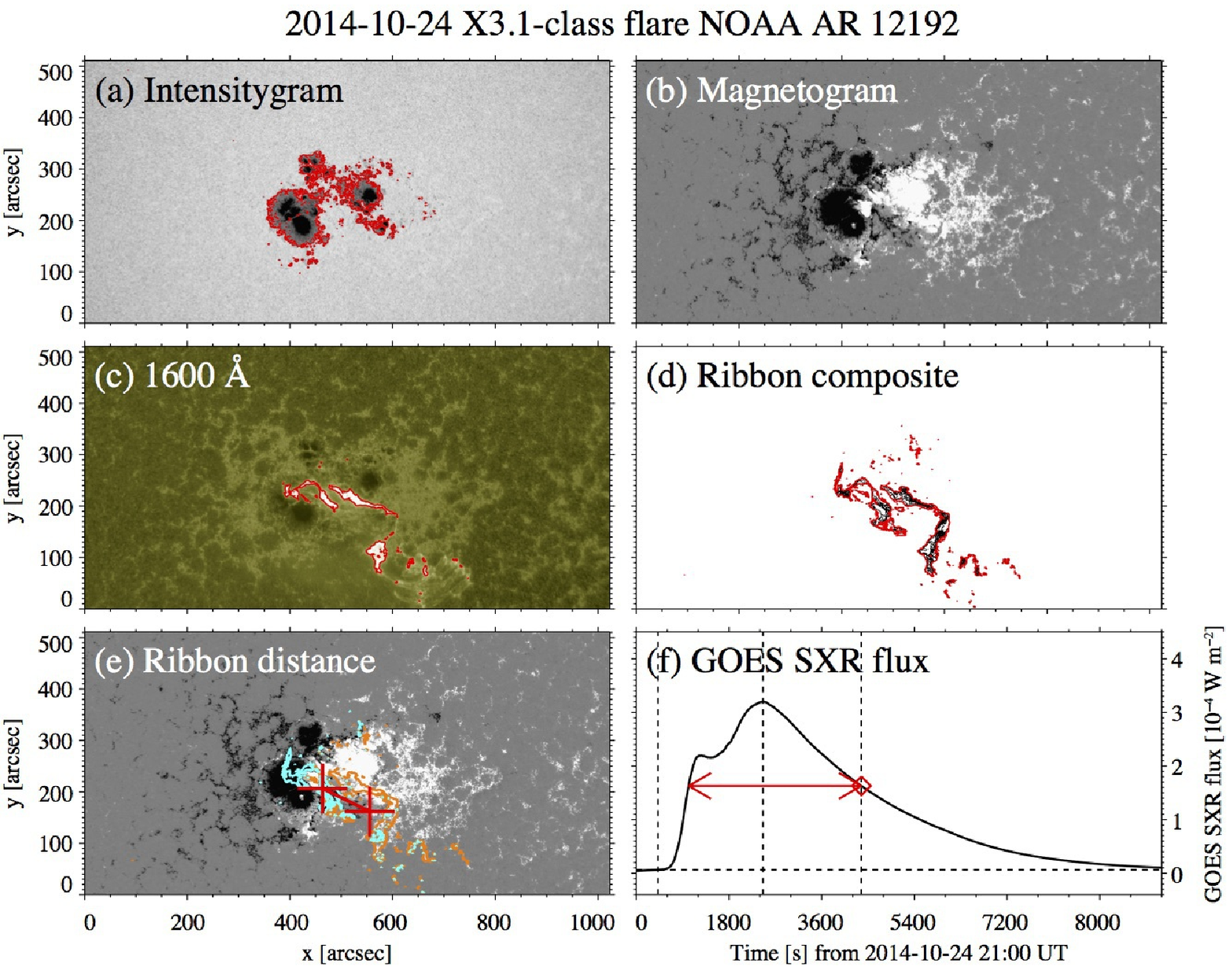}
  \end{center}
  \caption{Sample flare data:
    X3.1-class event
    in NOAA AR 12192
    (event \#37).
    (a) HMI intensitygram
    and (b) magnetogram (saturating at $\pm 400\ {\rm G}$),
    both taken at 2014-10-24 21:00 UT.
    Red contour in panel (a)
    surrounds the umbrae and penumbrae,
    defining the spot area.
    (c) AIA 1600 {\AA} image
    at 21:20 UT
    with red contour
    defining the flare ribbon
    (intensity of $\ge 40\sigma$ above the mean)
    in this frame.
    (d) Detected flare ribbons
    from some selected frames
    are overlaid (black).
    The red contour
    outlining these ribbons
    indicates the ribbon composite.
    (e) Composite ribbons
    in the positive (orange)
    and negative (turquoise) polarities
    plotted over the magnetogram (b).
    Red ``$+$'' signs show
    the area-weighted
    centroids of the two ribbons.
    A red straight line
    connects the two centroids,
    indicating the ribbon distance.
    (f) GOES SXR 1 -- 8\ {\AA} flux (solid curve).
    Three vertical dashed lines show
    (from left to right)
    the GOES start (21:07 UT),    
    peak (21:47 UT),
    and end (22:13 UT) times.
    The horizontal dashed line
    indicates the background level,
    which is the flux
    measured at the GOES start time.
    After subtracting this background level
    from the light curve,
    FWHM time is measured (red arrow),
    while the flux and its time derivative
    at the GOES end time (red diamond)
    are used for measuring the $e$-folding decay time.
    }
  \label{fig:sample}
\end{figure}

\begin{figure}
  \begin{center}
    \includegraphics[width=70mm]{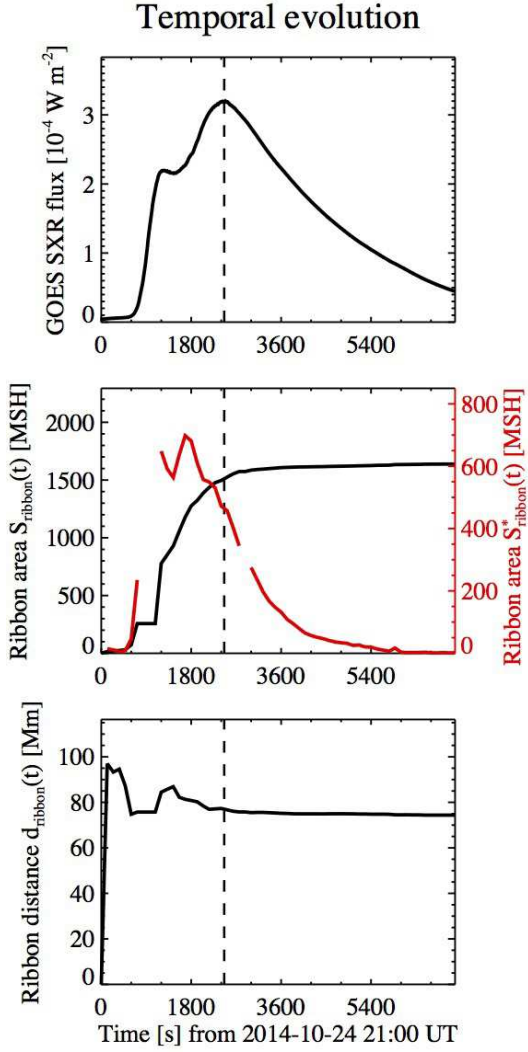}
  \end{center}
  \caption{
    Temporal evolution
    of the X3.1-class flare
    (event \#37).
    (Top) GOES SXR 1 -- 8\ {\AA} flux.
    (Middle) The area of the ribbon composite,
    $S_{\rm ribbon}(t)$ (black),
    which is calculated
    from the AIA 1600 {\AA} images
    until each moment, $t$,
    and the actual area of the ribbon
    at each moment,
    $S^{\ast}_{\rm ribbon}(t)$ (red).
    The periods of blank $S^{\ast}_{\rm ribbon}(t)$
    indicate the saturation
    in the 1600\ {\AA} images.
    (Bottom) Ribbon distance,
    $d_{\rm ribbon}(t)$,
    which is calculated
    from the ribbon composite
    at each moment $S_{\rm ribbon}(t)$.
    In all panels,
    the GOES peak time
    is indicated with vertical dashed line,
    which separates the impulsive phase
    and the gradual (decay) phase.
  }
  \label{fig:evolution}
\end{figure}

\begin{figure}
  \begin{center}
    \includegraphics{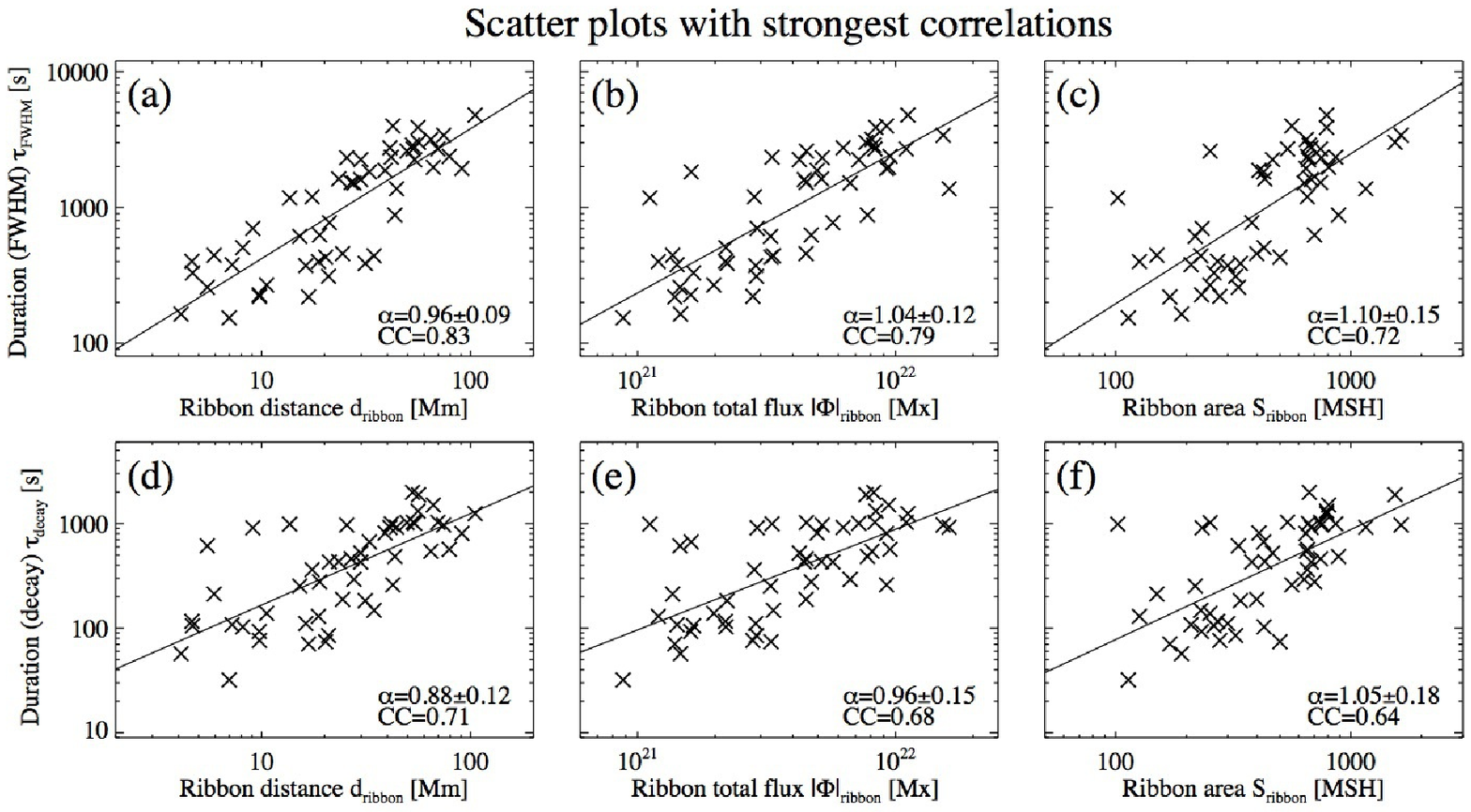}
  \end{center}
  \caption{Scatter plots
    with strongest correlations
    ($|CC|\ge 0.6$).
    In each panel,
    a straight line
    shows the result of a linear fitting
    to the log-log plots,
    while power-law index $\alpha$
    and correlation coefficient $CC$
    are shown at the bottom right.
    \label{fig:param}}
\end{figure}

\begin{figure}
  \begin{center}
    \includegraphics{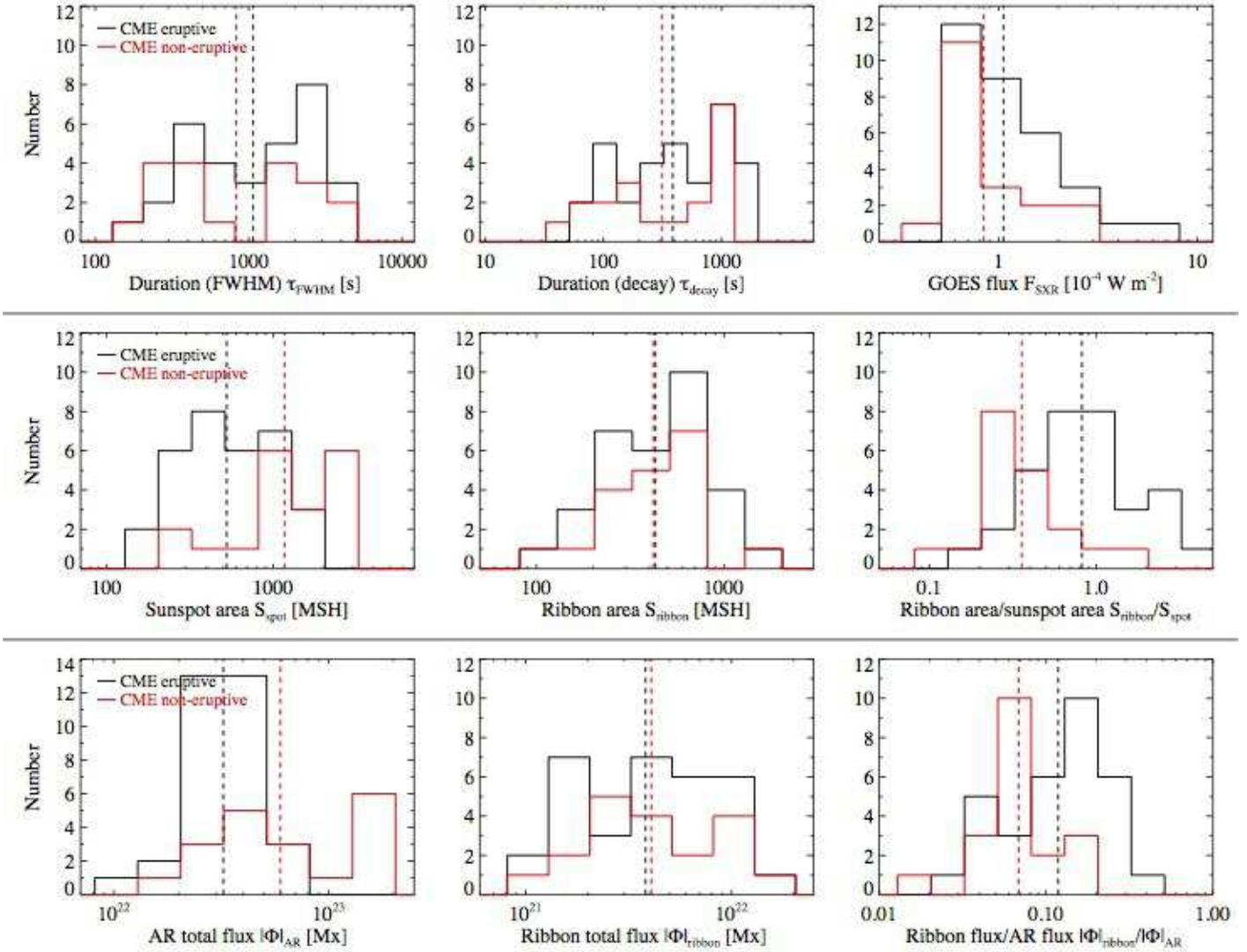}
  \end{center}
  \caption{Comparison of the histograms
    for CME eruptive (black)
    and non-eruptive (red) events:
    total event numbers
    are 32 and 19,
    respectively.
    Dashed vertical lines
    indicate
    the means of the log values.
    (Top) Histograms
    of FWHM duration $\tau_{\rm FWHM}$,
    decay time $\tau_{\rm decay}$,
    and GOES peak flux $F_{\rm SXR}$.
    (Middle) Histograms 
    of spot area $S_{\rm spot}$,
    ribbon area $S_{\rm ribbon}$,
    an their ratio $S_{\rm ribbon}/S_{\rm spot}$.
    (Bottom) Histograms of
    AR total magnetic flux $|\Phi|_{\rm AR}$,
    ribbon flux $|\Phi|_{\rm ribbon}$,
    and their ratio $|\Phi|_{\rm ribbon}/|\Phi|_{\rm AR}$.
    \label{fig:cmeyn}}
\end{figure}

One of the clear differences
is seen in the spot area, $S_{\rm spot}$.
In the second row of Figure \ref{fig:cmeyn},
distributions of $S_{\rm spot}$ show
a large discrepancy.
Here,
the non-eruptive events
have larger spot areas.
The log averages
are 526 MSH for eruptive
and 1171 MSH for non-eruptive
(difference $=76$\%),
and the spot areas
of the eruptive cases are
significantly smaller
than the non-eruptive
at the 99.5\% confidence level
(see Appendix \ref{app:student}).
As one might expect,
the latter value is to some extent
influenced by the six non-eruptive events
from the cycle's largest spot group,
AR 12192 (Figure \ref{fig:sample}).
However,
even without these flares,
the log-mean spot area
of the remaining 13 events
is still 801 MSH
(difference $=41$\%)
and the distribution difference
is significant
at 95\% confidence.
On the other hand,
the distributions and thus the mean values
of the ribbon area, $S_{\rm ribbon}$,
are similar
for the eruptive and non-eruptive:
the log means are
432 and 419 MSH,
respectively
(difference $=3.0$\%).

As a result of the differences
in $S_{\rm spot}$ and $S_{\rm ribbon}$,
the area ratios,
$S_{\rm ribbon}/S_{\rm spot}$,
also show a difference
in the distributions
with log-mean values
of 0.82 and 0.36
for eruptive and non-eruptive,
respectively
(difference $=79$\%).
The threshold
dividing the two regimes
is about 0.5.
This clear difference
may indicate that
what determines the CME productivity
is the relative structural relation
between the magnetic fields
of the flaring region
(sheared PIL, flare ribbons, flare arcades, etc.)
and those of the entire AR.

Tendencies similar to
those of the areas
($S_{\rm spot}$ and $S_{\rm ribbon}$)
are seen
for the total magnetic flux
(bottom row of Figure \ref{fig:cmeyn}).
The log-mean values
of $|\Phi|_{\rm AR}$
for the eruptive and non-eruptive
are $3.2\times 10^{22}$
and $6.0\times 10^{22}\ {\rm Mx}$
(difference $=59$\%;
significant at 99.5\% confidence),
respectively,
while those
of $|\Phi|_{\rm ribbon}$
are $3.8\times 10^{21}$
and $4.1\times 10^{21}\ {\rm Mx}$
(6.5\%),
respectively.
And thus the log means
of $|\Phi|_{\rm ribbon}/|\Phi|_{\rm AR}$
are 0.12 and 0.07
(53\%),
respectively.

The other three parameters,
$|B|_{\rm AR}$,
$|B_{\rm ribbon}|$,
and $d_{\rm ribbon}$,
are not very different
between the two cases:
the differences are
4.8\%, 9.5\%, and 7.7\%,
respectively.

Previous findings
of our earlier report, \citet{har16},
and of event studies
introduced in Section \ref{sec:introduction}
are confirmed
by the present
comprehensive
survey:
the present work
covers all on-disk flare events
over a six-year period
including the cycle maximum
without selection bias
and extends the on-disk sample
of \citet{har16}
with the GOES peak brightness
reaching down to M5 level,
which is
one virtue
of this study\footnote{For example,
20 on-disk flares
from the whole 42 X-class events
were used in the plot
for the flare duration versus the spot area
in \citet[][Figure 5]{har16}.
In the present work,
the sample number
of the on-disk events
is expanded to 51,
i.e., by a factor of 2.5,
which contains the previous 20 flares.}.

\begin{figure}
  \begin{center}
    \includegraphics{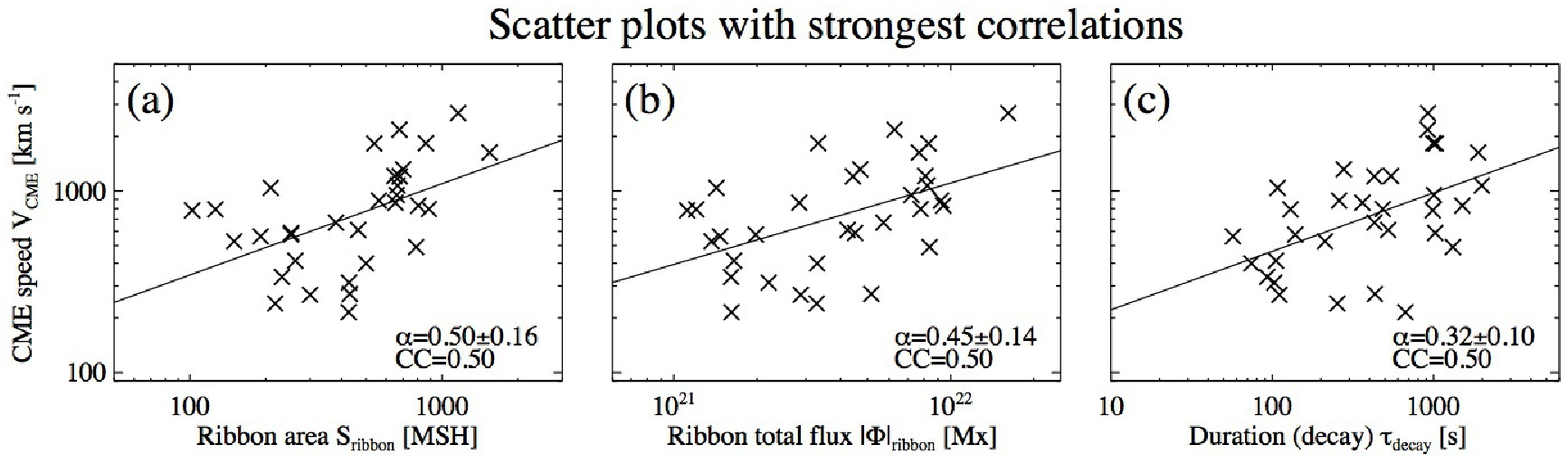}
  \end{center}
  \caption{Scatter plots
    with strongest correlations
    for the CME speeds $V_{\rm CME}$.
    In each panel,
    a straight line
    shows the result of a linear fitting
    to the log-log plots,
    while power-law index $\alpha$
    and correlation coefficient $CC$
    are shown at the bottom right.
    \label{fig:param4_2}}
\end{figure}

In order to find the parameters
that control the CME speed,
we made scatter plots
of $V_{\rm CME}$,
similar to those in Section \ref{sec:goes}
but this time
also of GOES parameters,
$\tau_{\rm FWHM}$, $\tau_{\rm decay}$, and $F_{\rm SXR}$.
The rightmost column
of Table \ref{tab:power} summarizes
the power-law indices, $\alpha$,
and their correlation coefficients, $CC$.
The largest value is $CC=0.5$
for the ribbon area
$S_{\rm ribbon}$,
ribbon flux
$|\Phi|_{\rm ribbon}$,
and GOES decay time
$\tau_{\rm decay}$,
which are shown in Figure \ref{fig:param4_2}.
Note that
for a sample number of 32,
any correlation over 0.45
is significant
at 99\% confidence.
In this study,
we only selected
the flares that occurred
within 45$^{\circ}$ from disk center,
which makes the $V_{\rm CME}$ values
rather uncertain.
Projection effects
due to non-radial motions
may also increase the scatter.
However,
even with such uncertainties,
the results show
significant higher correlation.

\section{Magnetic Patterns of Flare Zones}\label{sec:classification}

\begin{figure}
  \begin{center}
    \includegraphics[width=170mm]{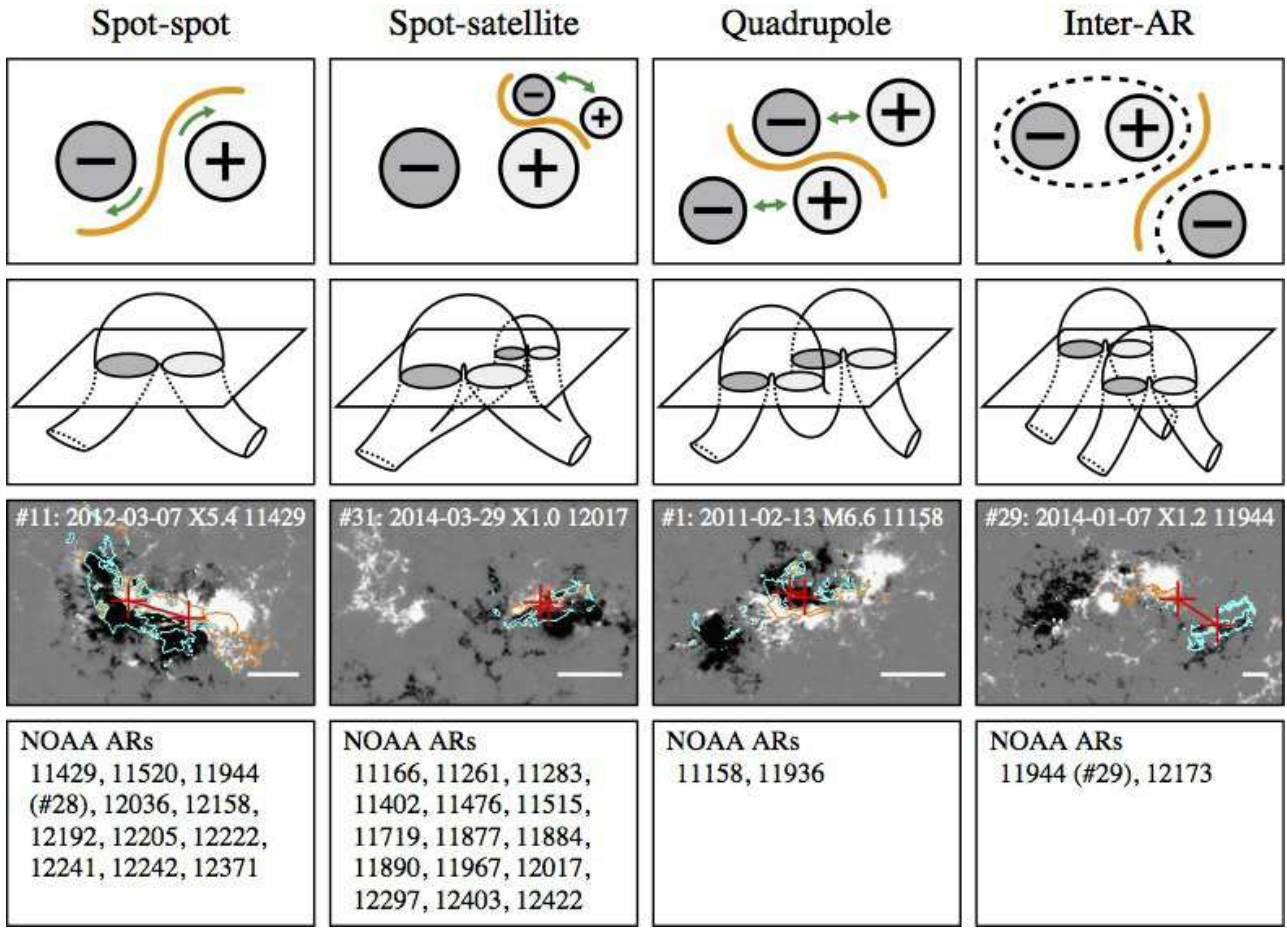}
  \end{center}
  \caption{Characterization
    of flaring ARs.
    (Top)
    Polarity distributions.
    Magnetic elements (spots)
    are indicated
    by circles
    with ``$+$'' and ``$-$'' signs.
    The PIL involved in the flare
    is shown with orange line,
    while proper motions of the polarities
    are indicated with green arrows.
    (Second)
    Possible three-dimensional structures
    of magnetic fields.
    Solar surface is indicated
    with a horizontal slice.
    (Third)
    Sample events.
    Event number, date,
    GOES class, and NOAA number,
    are shown at the top.
    Contours and ``$+$'' signs are
    identical to those in Figure \ref{fig:sample}(e).
    White line at the bottom right
    indicates the length of $50\arcsec$.
    (Bottom)
    NOAA numbers of the corresponding ARs.
    Event numbers (Table \ref{tab:flares1})
    are also shown
    for AR 11944
    to distinguish its two flare events.
    \label{fig:classification}}
\end{figure}

In flaring ARs,
sheared magnetic structures
responsible to the flare productions,
such as sheared PILs,
are probably created by
the large-scale flux emergence
and the (resultant) relative motions
of the sunspots
\citep[e.g.,][]{kur89}.
Besides,
the geometrical relationship
between the sheared PILs
and the entire AR
may determine the characteristics
of the flare events.
Therefore,
in this section,
we focus on
the creation of sheared PILs
in the entire ARs
and investigate the flare production
in different types of ARs.
The details of this
characterization
are summarized
in Figure \ref{fig:classification}.

The first
characterization
is the ``spot-spot'' group,
in which a large, long sheared PIL extends
across the entire AR
between the two major polarities
or between the two clusters of sunspots
of opposite polarities.
Such ARs may naturally harbor
large flare ribbons.
Among the 11 ARs (21 events)
that belong to this category
(see bottom of Figure \ref{fig:classification}),
NOAA AR 11429 produced
the strongest (X5.4-class) flare
so far in this solar cycle.
Based on the numerical simulation
of flux emergence,
\citet{tak15} suggested the possibility that
AR 11429 was created by
an emergence of
a tightly-twisted, kink-unstable flux tube
\citep[see also, e.g.,][]{tan91,lin96,fan98}.
The spot-spot may also be created
by many episodes of flux emergence.

The second group
is that of the
``spot-satellite''.
25 events from 15 ARs belong to this category.
Here,
newly emerging,
often minor, magnetic flux
appears just next to
one of the pre-existing main polarities
and creates a compact PIL
between the main and satellite spots.
Such a close emergence of satellite spots
hints that
the satellite spots are connected
to the main polarity
below the surface
as a parasite tube,
like illustrated
in Figure \ref{fig:classification}.
Or perhaps the satellite spots
are from
an independent minor flux tube,
which is floating
in the convection zone
and trapped by the main tube
that rises through the interior.
The ``best-observed'' X1.0-class flare
of AR 12017
(event \#31: \citealp{kle15})
falls into this category.

Then,
the ``quadrupole'' group follows
these two majorities
(three events, two ARs).
In this group,
two opposite polarities
from different emerging bipoles
collide each other,
show shear motion,
and create a sheared PIL
in between.
By comparing flux emergence simulation
and observational data,
\citet{tor14a} obtained a suggestion that
AR 11158 is created from a single flux tube
that emerges at the two locations
\citep{fang15}.

The last group,
``inter-AR'',
is of the flares produced
on the PIL formed
between two apparently independent ARs
(two events from different AR pairs).
The clearest example
is the X1.2-class flare
(event \#29: \citealp{moe15}),
which occurred between AR 11944
and the decayed AR 11943.
This category resembles
the quadrupole events.
However,
we here divide
these two groups
by whether the flare occurred
between the polarities
that belong to a single NOAA-numbered AR,
or between the polarities
of independent ARs with different NOAA numbers,
since this categorization may imply
whether a mutual (subsurface) magnetic connectivity
exists or not.
In fact,
neither of the inter-AR events have
a $\delta$-configuration
at the flaring site.
And thus,
this group reminds us of
the eruption of a quiescent filament,
which is created in the quiet Sun
between extended AR remnants,
probably with the support
of shear flows
caused by the differential rotation
\citep{mac10}.
Perhaps,
the inter-AR events occupy
an intermediate position
between the flares from ARs
and those of quiescent filament eruptions.

\begin{figure}
  \begin{center}
    \includegraphics{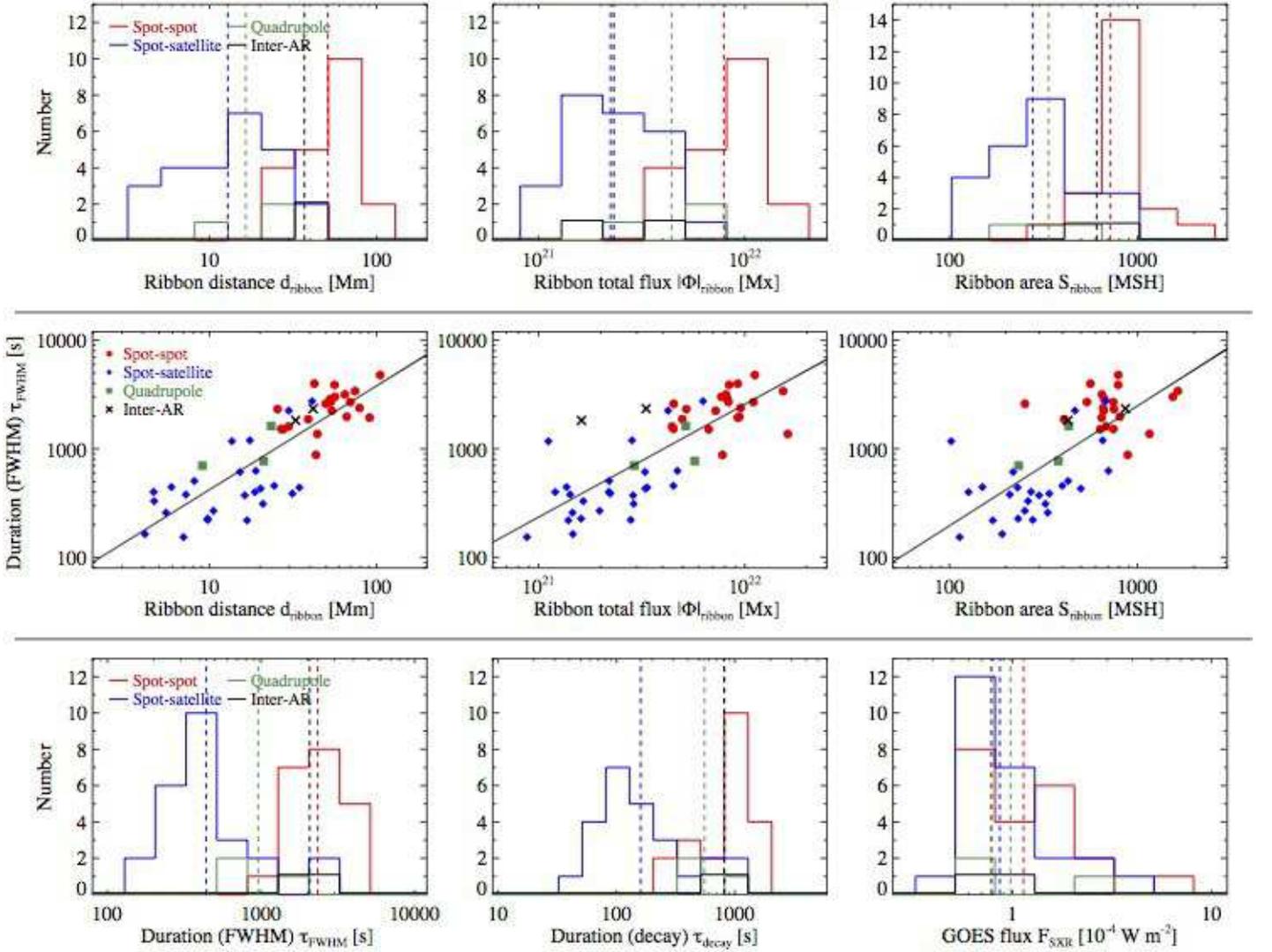}
  \end{center}
  \caption{Statistical trends
    of the flares
    of different
    magnetic patterns.
    (Top) Histograms
    for the ribbon distance $d_{\rm ribbon}$,
    ribbon flux $|\Phi|_{\rm ribbon}$,
    and ribbon area $S_{\rm ribbon}$.
    Colors represent the
    patterns:
    spot-spot (red),
    spot-satellite (blue),
    quadrupole (green),
    and inter-AR (black).
    (Middle) Scatter plots
    of the flare duration $\tau_{\rm FWHM}$
    versus $d_{\rm ribbon}$,
    $|\Phi|_{\rm ribbon}$,
    and $S_{\rm ribbon}$,
    i.e.,
    the same as Figures \ref{fig:param}(a--c)
    but with different symbols.
    Straight lines show
    the fitting results.
    (Bottom) Histograms
    of the FWHM duration $\tau_{\rm FWHM}$,
    $e$-folding decay time $\tau_{\rm decay}$,
    and GOES peak flux $F_{\rm SXR}$.
    \label{fig:classification_result}}
\end{figure}

In reviewing these four patterns
identified in flaring regions,
we note that
the spot-spot group may possess
larger flare ribbons
since the flares of this group
are likely to occur
above the extended sheared PILs
across the entire ARs.
Conversely,
the spot-satellite events
are expected
to have smaller ribbons.
The top row
of Figure \ref{fig:classification_result}
clearly shows the above trends.
Here, the spot-spot events have
larger ribbon distance, ribbon flux, and ribbon area,
while the spot-satellite flares
have smaller values.
The log averages of the above parameters
for the spot-spot and spot-satellite are
$d_{\rm ribbon}=51.0$
and 12.9 Mm (difference $=119$\%),
$|\Phi|_{\rm ribbon}=7.9\times 10^{21}$
and $2.3\times 10^{21}\ {\rm Mx}$
(112\%),
$S_{\rm ribbon}=715$
and 277 MSH
(88\%).
The quadrupole and inter-AR values
generally sit
between the two major groups.

Then,
through the statistical relations
(\ref{eq:tfwhm_dribbon}), (\ref{eq:tfwhm_pribbon}),
and (\ref{eq:tfwhm_sribbon}),
the spot-spot events
have longer GOES durations,
and the spot-satellite ones
are shorter
(middle row of Figure \ref{fig:classification_result}).
Again,
the quadrupole and inter-AR events
are in the intermediate positions.
As seen from the bottom row
of this figure,
the spot-spot events have
the FWHM durations
of $\gtrsim 1000\ {\rm s}$,
the spot-satellite $\lesssim 1000\ {\rm s}$.

Similar trends are obtained
for the $e$-folding decay time
through relations
(\ref{eq:tdecay_dribbon}), (\ref{eq:tdecay_pribbon}),
and (\ref{eq:tdecay_sribbon}).
The critical value
dividing the two regimes
is $\tau_{\rm decay}\sim 200\ {\rm s}$.
However,
the GOES peak flux
does not show
a prominent contrast
between the distributions
of the spot-spot and spot-satellite:
log mean values are $1.1\times 10^{-4}$
and $0.86\times 10^{-4}\ {\rm W\ m}^{-2}$,
respectively
(difference $=27$\%).

These results lead us to the conclusion that
the structural differences
of the flaring ARs
determine the size of the sheared PILs
and thus of the flare ribbons,
which dictate the flare durations.
On the other hand,
the GOES flux has much weaker relation
with the structural differences,
which we hypothesize to reflect
that other factors than only geometry
are involved
in setting the total energy
and intensity profile
of a flare.

The fractions
of the CME-eruptive events
for the spot-spot, spot-satellite, quadrupole, and inter-AR events
are 57\% (12 in 21 events),
64\% $(=16/25)$,
67\% $(=2/3)$,
and 100\% $(=2/2)$,
respectively.
Therefore,
the spot-spot events are
less likely to be CME-eruptive
than the spot-satellite events.
This result is well in line with
the discussions
in the previous sections
that a strong overlying arcade,
which is likely to exist
in a spot-spot AR,
prohibits the CME eruption.
However,
because of the small sample numbers,
it is difficult to make
any firm conclusion
on the quadrupole and inter-AR events.

\section{Possibilities of Historical Data Analysis}\label{sec:historical}

\begin{figure}
  \begin{center}
    \includegraphics{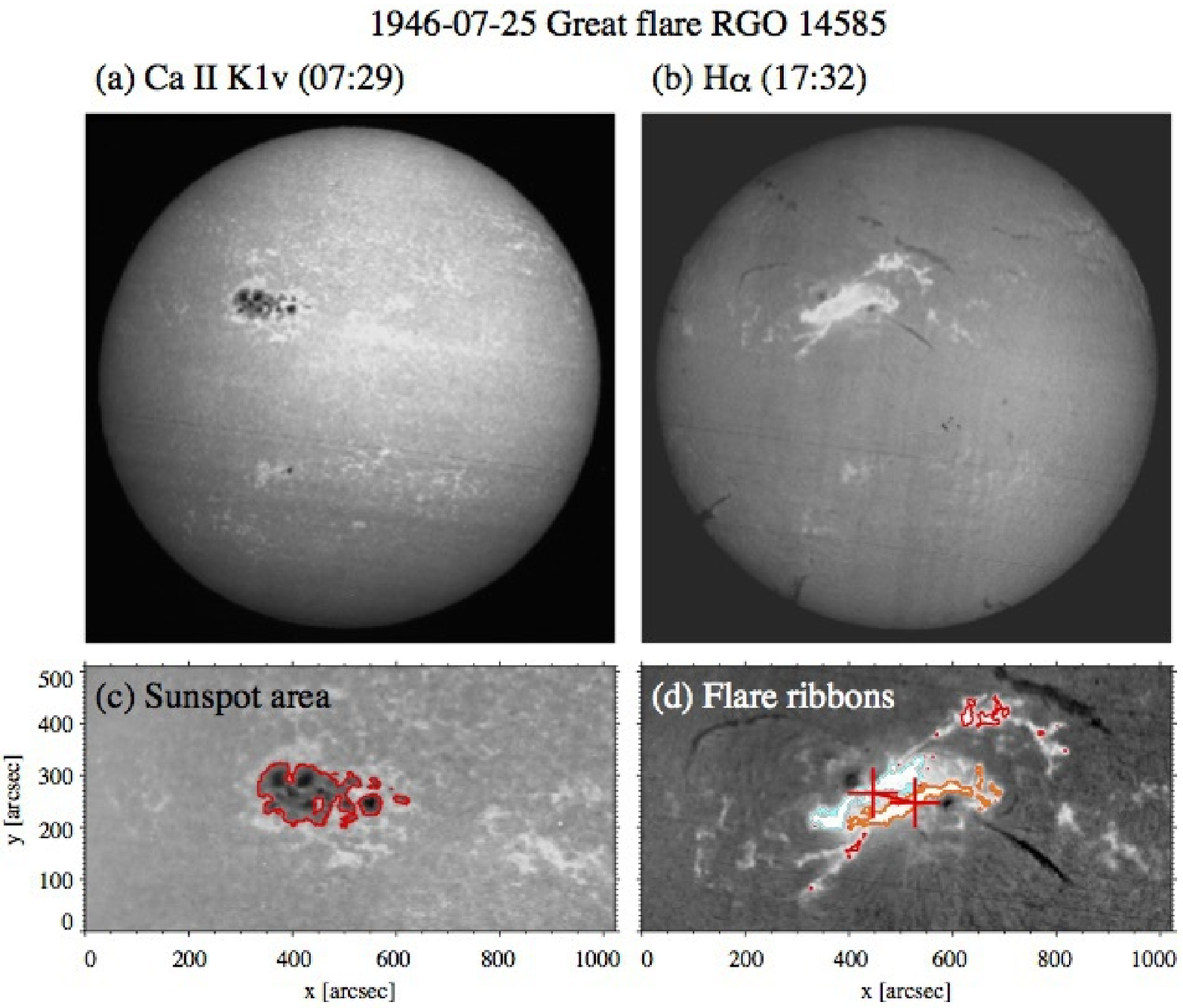}
  \end{center}
  \caption{Great flare event
    SOL1946-07-25
    in RGO 14585,
    observed by the Meudon spectroheliopraph.
    (a) \ion{Ca}{2} K1v and
    (b) H$\alpha$ full-disk images.
    (c) Cutout of (a) showing the spot area
    (red contour).
    Threshold is set to be 90\%
    of the mean quiet-Sun intensity
    after the background trend is subtracted.
    (d) Cutout of (b) showing the ribbon area and distance.
    Red, orange, and turquoise contours
    indicate the brightest regions
    in this image,
    i.e., ribbon area.
    Threshold is set to be
    280\% of the mean quiet-Sun intensity
    after the background trend is subtracted.
    From the two largest patch groups
    represented by orange and turquoise,
    we measured the distance
    between the two centroids,
    i.e., ribbon distance
    (red ``$+$'' signs and a straight line).}
  \label{fig:historical}
\end{figure}

Figure \ref{fig:historical} shows
perhaps the largest-ever imaged sunspot-related flare ribbons.
This sunspot group,
numbered 14585
by the Royal Greenwich Observatory (RGO)
and 8129
by the Mount Wilson Observatory,
produced a great flare
on 1946 July 25
(flare importance $3+$:
\citealt{ell49});
in modern usage SOL1946-07-25.
In the list of
Sunspot Groups with Largest Areas
maintained at
NAOJ\footnote{\url{http://solarwww.mtk.nao.ac.jp/en/bigspots.html}},
RGO 14585 ranks fourth.
According to \citet{dod49}
and RGO reports,
this region had a spot size
of 4279 MSH that day
with $\beta\gamma$-configuration.
\citet{ell46}
observed in H$\alpha$ and many other lines and
reported that
the great flare
continued for several hours.
It started
before 16:15 UT
and reached
its maximum intensity
around 16:30 UT.
By 17:30 UT,
a bright emission
had increased to
2500 MSH in area,
accompanied by
a filament
of 550 Mm in length.
His observation continued
at least
until 18:10 UT.
The flare caused
a great geomagnetic storm
26.5 hours later,
and even triggered
a ground level enhancement
(GLE: \citealt{for46,neh48}).
This region repeatedly
produced flare eruptions
\citep{dod49}.

As is seen from Figure \ref{fig:historical},
this region is composed
of a number of spots,
i.e., highly fragmented.
On the other hand,
it exhibits
a giant flare ribbon
that extends over
the entire region.
In fact,
our measurement
of the spot size
in \ion{Ca}{2} K1v,
$S_{\rm spot}$,
and ribbon size in H$\alpha$, $S_{\rm ribbon}$,
are 4200 and 3570 MSH
(projection corrected values),
respectively.
Here,
$S_{\rm ribbon}$
might be underestimated
because
the ribbon possibly
expanded more
in the later phase,
and thus
the area ratio,
$S_{\rm ribbon}/S_{\rm spot}$,
is {\it at least} 85\%,
which indicates that
a considerable fraction
of RGO 14585
was involved
in the flare production.

\begin{figure}
  \begin{center}
    \includegraphics[width=85mm]{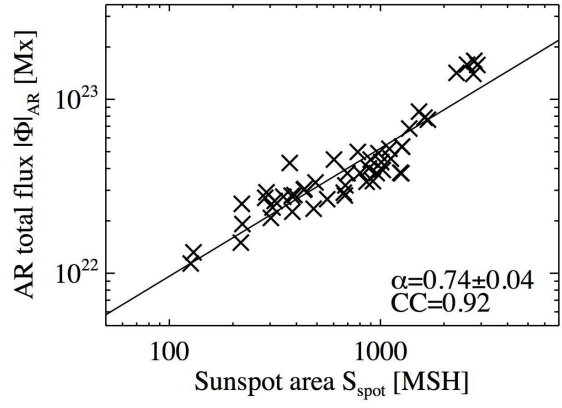}
  \end{center}
  \caption{
    Scatter plot
    of AR total flux $|\Phi|_{\rm AR}$
    versus spot area $S_{\rm spot}$
    for the 51 target events.
    Black straight line is
    the result of linear fitting
    to the log-log plot.
    \label{fig:flux}}
\end{figure}

We can place this region
in the context
of our present sample
through the relationship
in Figure \ref{fig:flux},
which shows
the scatter plot
of AR total flux $|\Phi|_{\rm AR}$
versus spot area $S_{\rm spot}$
for the 51 $\ge$M5-class events
that we analyzed
in Section \ref{sec:results}.
Note that $S_{\rm spot}$ indicates
the total area of umbrae and penumbrae,
i.e., the sunspot area,
rather than
the area of the entire AR.
The linear fitting to this log-log plot
provides the relation of
\begin{eqnarray}
  \log{|\Phi|_{\rm AR}}=(0.74\pm 0.04)\log{S_{\rm spot}}+(20.5\pm 0.13).
  \nonumber\\
  \label{eq:par_sar}
\end{eqnarray}
Using this equation,
the measured spot area of 4200 MSH
on July 25
yields a flux of
$1.5\times 10^{23}\ {\rm Mx}$,
which is comparable
to the maximum
of our $\ge$M5 data set,
(1.4 -- $1.7)\times 10^{23}\ {\rm Mx}$
of AR 12192.
Although this region appeared
before the $\delta$ classification
was introduced
by \citet{kue60},
this region is likely to possess
a $\delta$-configuration
since the long flare ribbons
lie in the middle of the spots
that share common penumbrae.
For the same reason,
we can categorize
this region
as ``spot-spot''.

In addition,
from the H$\alpha$ image,
we estimated
the ribbon distance $d_{\rm ribbon}$.
We here took
the two largest ribbon groups
and measured the distance
between the centroids:
see Figure \ref{fig:historical}(d).
Through equation (\ref{eq:tfwhm_dribbon}),
the obtained value,
$d_{\rm ribbon}=62\ {\rm Mm}$,
which should also be considered
as a lower limit,
yields the FWHM duration,
$\tau_{\rm FWHM}$, of 2400 s.
The actual duration of the flare event
is not clear
but may be a few times
of this value,
say, a few hours.
In fact,
the observations revealed that
the flare continued
at least for 110 minutes
(\citealt{ell46}:
observed mainly in H$\alpha$).

Furthermore,
the large area ratio
of this event,
$S_{\rm ribbon}/S_{\rm spot}\ge 85$\%,
implies
the occurrence of a CME
(see, e.g., Figure \ref{fig:cmeyn}).
In fact,
the great flare
caused a geomagnetic storm
after 26.5 hours
and even a GLE
\citep{for46},
which suggests the existence
of a severe disturbance
such as a fast CME.

\section{Discussion}\label{sec:discussion}

\subsection{Interpretation of the Obtained Relations}\label{sec:interpretation}

In this study,
we have conducted
a statistical analysis
of 51 solar flares
with GOES magnitude $\ge$M5
emanating from 29 ARs,
and have obtained
six high-correlation ($|CC|\ge 0.64$) empirical relations,
(\ref{eq:tfwhm_dribbon}) -- (\ref{eq:tdecay_sribbon}).
They indicate that
the durations of the GOES light curves
(FWHM duration $\tau_{\rm FWHM}$
and $e$-folding decay time $\tau_{\rm decay}$)
correlate linearly
with the flare parameters
(ribbon distance $d_{\rm ribbon}$,
ribbon total flux $|\Phi|_{\rm ribbon}$,
and ribbon area $S_{\rm ribbon}$).
If we use $\tau_{\rm flare}$
to simply denote
the flare duration,
they can be
characterized by
\begin{eqnarray}
  \tau_{\rm flare}\propto d_{\rm ribbon},
  \label{eq:tflare_dribbon}
\end{eqnarray}
\begin{eqnarray}
  \tau_{\rm flare}\propto |\Phi|_{\rm ribbon},
  \label{eq:tflare_pribbon}
\end{eqnarray}
and
\begin{eqnarray}
  \tau_{\rm flare}\propto S_{\rm ribbon}.
  \label{eq:tflare_sribbon}
\end{eqnarray}
In this section,
we discuss the physical interpretations
of these relations.

\begin{figure}
  \begin{center}
    \includegraphics[width=80mm]{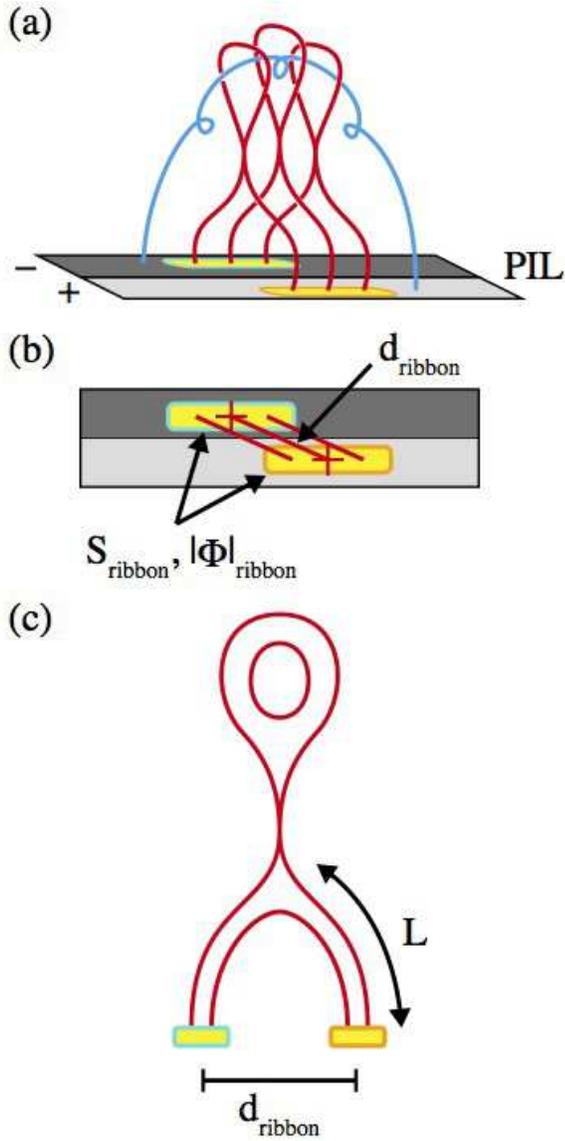}
  \end{center}
  \caption{
    Schematic illustrations
    of the standard flare model.
    (a) Filament (cyan), or flux rope,
    above the PIL
    between positive ($+$) and negative ($-$) polarities erupts
    and overlying coronal magnetic fields (red)
    reconnect under the ascending filament.
    As a result,
    flare ribbons
    (yellow regions
    outlines by orange and turquoise lines)
    are created in the chromosphere.
    (b) Top view of (a).
    Ribbon distance $d_{\rm ribbon}$,
    ribbon area $S_{\rm ribbon}$,
    and ribbon total flux $|\Phi|_{\rm ribbon}$
    are indicated.
    Red ``$+$'' signs show
    the centroids of the ribbons.
    (c) Side view of (a).
    Half length of
    the reconnected (post-flare) loops $L$
    is indicated
    along with the ribbon distance $d_{\rm ribbon}$.
    \label{fig:model}}
\end{figure}

First,
what do the flare parameters,
$d_{\rm ribbon}$, $|\Phi|_{\rm ribbon}$,
and $S_{\rm ribbon}$,
mean?
As we mentioned
in Section \ref{sec:introduction},
in the standard (CSHKP) flare model,
the flare ribbons are caused by
coronal magnetic energy released into
high-energy electrons
and thermal conduction.
Therefore,
we can assume that
the ribbons are
the footpoints of
newly reconnected post-flare loops.
Figure \ref{fig:model} shows
schematic illustrations
of the standard model.
As the filament erupts,
overlying coronal fields reconnect
under the filament,
and the post-flare loops
and flare ribbons are created.
From Figures \ref{fig:model}(b) and (c),
it is seen that
the distance between the two centroids
of the ribbon composite,
$d_{\rm ribbon}$,
indicate the footpoint separation
between the representative post-flare loop.
If the loop configuration does not
differ
much
for different flare events,
the loop half length, $L$,
would be proportional to $d_{\rm ribbon}$,
i.e., $L\propto d_{\rm ribbon}$.
Meanwhile,
$|\Phi|_{\rm ribbon}$ indicates
the total magnetic flux
in the ribbon composites,
identifiable with the flux
involved
in the flare reconnection,
whereas $S_{\rm ribbon}$ is
the total area of the composite.

For explaining
the first relation,
$\tau_{\rm flare}\propto d_{\rm ribbon}$,
we here simply assume that
the duration of the flares,
especially the evolutionary phase
when the ribbon expansion occurs
(see Section \ref{sec:sample}),
is comparable to the reconnection time scale,
i.e., $\tau_{\rm flare}\sim \tau_{\rm rec}$.
This time scale is roughly estimated as
$\tau_{\rm rec}\sim L/V_{\rm in}$,
where $V_{\rm in}$ is the velocity
of pre-reconnection magnetic fields
flowing into the electric current sheet,
and this relation is rewritten as
$\tau_{\rm rec}\sim\tau_{\rm A}/M_{\rm A}$,
where $\tau_{\rm A}\equiv L/V_{\rm A}$
is the Alfv\'{e}n transit time over the loop,
$V_{\rm A}$ the Alfv\'{e}n velocity,
and $M_{\rm A}=V_{\rm in}/V_{\rm A}$
the Alfv\'{e}n Mach number.
If we
assume
from Figure \ref{fig:model}(c) that
$L\sim d_{\rm ribbon}$,
we find
the proportionality
$\tau_{\rm flare}\sim d_{\rm ribbon}/(V_{\rm A}M_{\rm A})$.

It is seen for example
from Figure \ref{fig:param}(a) that
$V_{\rm A}M_{\rm A}=20$ -- $30\ {\rm km\ s}^{-1}$,
and applying
$M_{\rm A}=0.01$ -- $0.1$,
the typical values
for the Petschek-type reconnection model
\citep{pet64}
obtained from
resistive-MHD
simulations
\citep[e.g.,][]{yok97,yok98},
one may find that
$V_{\rm A}$ ranges
from a few 100 to a few $1000\ {\rm km\ s}^{-1}$.
Such values can be consistent with
the Alfv\'{e}n speed
inferred observationally for the solar corona
\citep[e.g.,][]{gop01},
though not the core of an active region,
indicating that
the above estimation is fairly plausible.

However,
because
the observed characteristics
in this study
such as the flare time scales
are the result of the superposition
of elementary flare loops,
and
because
each flare loop experiences
different stages
of thermal processes
after the reconnection
(i.e., the chromospheric evaporation,
conductive cooling, and radiative cooling),
which may have different time scales
\citep{rea07},
it is of high importance
to conduct MHD simulations
of flare reconnection and post-flare loops
including thermodynamic processes
in order to explore
the essential physics
involved
in the loop.

Relation (\ref{eq:tflare_pribbon}),
$\tau_{\rm flare}\propto |\Phi|_{\rm ribbon}$,
may be easier
to understand:
as more magnetic flux is involved,
the reconnection processes continue longer.
If the reconnection rate is comparable
for various events,
the ribbon area also
could
have a linear proportion,
i.e., relation (\ref{eq:tflare_sribbon}),
$\tau_{\rm flare}\propto S_{\rm ribbon}$.

In any case,
the clear correlations
between the flare duration
and flare parameters
(ribbon distance, magnetic flux, and area),
especially those of
$\tau_{\rm flare}\propto d_{\rm ribbon}$,
strongly point to
the physical connections
underlying them.
For example,
recently it has been suggested that
the impulsive events
with shorter ribbon distance,
$d_{\rm ribbon}$,
yield
more intense
white-light flares
\citep{wat16}.
This may imply that the loop physics
of compact coronal loops, with smaller $L$,
corresponds to more intense energy release
deeper in the photosphere.
We may utilize
this relation
in the opposite manner.
The observation of flare durations
may allow us to investigate
the physical states
of the reconnected loops,
such as those
of unresolved stellar flares
\citep[e.g.,][]{mul06}.

\subsection{Emergence, Flares, and Superflares}

One of the important lessons
we have learned
is that major flares are produced
from various types of ARs.
X-class events are produced
not only from the classical $\delta$-spots
such as those
classified by \citet{zir87},
or ``spot-spot'', ``spot-satellite'',
and ``quadrupole'' in this study,
but even from the PILs
between separated, independent ARs
with no $\delta$-configurations,
i.e., ``inter-AR'',
like the X1.2 event from ARs 11944 and 11943
(event \#29).

Also,
the fraction of the region
that is involved
in the flare reconnection
in a single AR
differs substantially.
The area of the ribbon composite
normalized by the spot area,
$S_{\rm ribbon}/S_{\rm spot}$,
ranges from 9.0 to 
300\%
for the analyzed $\ge$M5 flares
(except for the three inter-AR events),
while the ribbon flux
normalized by AR flux,
$|\Phi|_{\rm ribbon}/|\Phi|_{\rm AR}$,
ranges from 1.6 to 43\%.

Therefore,
we need a systematic survey
using flux emergence simulations
to model these types of ARs
\citep{tor14a,fang15,tak15,cha16}
and investigate their formation processes
as well as the storage of magnetic energy
(amount, place, etc.).
In Figure \ref{fig:cmeyn},
we found that
$S_{\rm ribbon}/S_{\rm spot}$
and $|\Phi|_{\rm ribbon}/|\Phi|_{\rm AR}$
are larger for the CME-eruptive events,
which may indicate
the importance
of the relative magnetic structure
of the flaring region and the entire AR.
Thus,
numerical experiments
on flux emergence
and flare AR formation
are necessary also
for the investigation
on the CME productions.

From the statistical analysis
of the stellar flares
obtained by {\it Kepler} space telescope,
\citet{mae12} suggested that
superflares with energy of
$10^{34}\ {\rm erg}$
occur once in 800 years
on the Sun-like stars
(slowly rotating G-type main sequence stars).
\citet{shi13} showed
through order-of-magnitude estimations that
in typical solar dynamo models,
it may be possible
to generate
a large sunspot
with a total flux of
$2\times 10^{23}\ {\rm Mx}$,
which accounts for
the flare of $10^{34}\ {\rm erg}$,
within one solar cycle period.
On the other hand,
\citet{aul13} argued that
superflares of $10^{34}\ {\rm erg}$
are unrealistic
for observed solar conditions
because of the fragmentation
of magnetic flux
in an AR:
all large sunspot groups
are highly fragmented,
i.e., composed
of many flux emergence events,
and thus
magnetic shear
tends to be localized.
Therefore,
only parts of the sunspots
might be involved
in the flare reconnection process
\citep[see also][]{sch12}.

However,
as we saw in Section \ref{sec:historical},
even one of the largest, highly fragmented sunspot groups
such as RGO 14585
could spout a flare eruption
leaving AR-sized, gigantic flare ribbons,
which may point to
the possibility that
even larger ARs
could occur
and cause a superflare.
The largest sunspot group
since the 19th century,
RGO 14886,
recorded a maximum spot area of 6132 MSH
on 1947 April 8
(see Figure 3 of \citealt{aul13}).
From equation (\ref{eq:par_sar}),
we estimate its total flux to be
$2.0\times 10^{23}\ {\rm Mx}$.
Therefore,
considering the two factors that
one of the largest ARs
produced the AR-scale eruption
and an AR of $2\times 10^{23}\ {\rm Mx}$
is likely to have existed,
we cannot completely rule out
the possibility that
an AR of $2\times 10^{23}\ {\rm Mx}$
produces AR-scale eruptions.
We will then estimate
the flare energy
in the next section.

\subsection{Estimation of Flare Energy}

The magnetic energy
that we discuss
in this section
is given using flare parameters as
\begin{eqnarray}
  &&E_{\rm mag}
  \sim \frac{B^{2}}{8\pi}\,V_{\rm mag}
  \sim \frac{\overline{|B|}_{\rm ribbon}^{2}}{4\pi}\,
  S_{\rm ribbon}\, d_{\rm ribbon}
  \nonumber \\
  \sim&& 4.3\times 10^{32}
  \left(
    \frac{\overline{|B|}_{\rm ribbon}}{325\ {\rm G}}
  \right)^{2}
  \left(
    \frac{S_{\rm ribbon}}{519\ {\rm MSH}}
  \right)
  \left(
    \frac{d_{\rm ribbon}}{32.9\ {\rm Mm}}
  \right)\ {\rm erg},
  \nonumber\\
  \label{eq:emag}
\end{eqnarray}
where $V_{\rm mag}$ is
the volume of magnetic fields
involved in the flare reconnection.
We assume here that
$V_{\rm mag}\sim 2S_{\rm ribbon}L$,
where $L$ is
the half length
of the reconnected loop
(Figure \ref{fig:model}),
and that $L\sim d_{\rm ribbon}$.
The parameters
used in this equation
for deriving
the typical value
are the means
from the 51 analyzed events,
and the estimated magnetic energy ranges
from $9.2\times 10^{30}$
to $4.4\times 10^{33}\ {\rm erg}$.

The magnetic energy (\ref{eq:emag}) may provide
better estimates for the flare energy,
$E_{\rm flare}\sim fE_{\rm mag}$,
where $f$ is the fraction
of the magnetic energy
that is released in the flare event,
compared to another expression
\citep[e.g.,][]{mae12,shi13,aul13}:
\begin{eqnarray}
  E_{\rm mag}
  &\sim& \frac{B^{2}}{8\pi}\,V_{\rm mag}
  \sim \frac{\overline{|B|}_{\rm AR}^{2}}{8\pi}\,
  S_{\rm spot}^{3/2}
  \nonumber \\
  &\sim& 2.9\times 10^{33}
  \left(
    \frac{\overline{|B|}_{\rm AR}}{688\ {\rm G}}
  \right)^{2}
  \left(
    \frac{S_{\rm spot}}{954\ {\rm MSH}}
  \right)^{3/2}\ {\rm erg}.
  \nonumber\\
\end{eqnarray}

For the great flare event
of RGO 14585
(SOL1946-07-25),
from equation (\ref{eq:emag})
with $\overline{|B|}_{\rm ribbon}\sim 384\ {\rm G}$
(mean of the spot-spot events),
$S_{\rm ribbon}\sim 3570\ {\rm MSH}$,
and $d_{\rm ribbon}\sim 62\ {\rm Mm}$,
the energy estimate becomes
$8\times 10^{33}\ {\rm erg}$.
If we suppose the situation that
the largest sunspot group
RGO 14886
($S_{\rm spot}=6132\ {\rm MSH}$ on 1947 April 8)
causes a whole-AR-scale eruption
like the 1946 event,
which may not be very unrealistic,
using the values of
$\overline{|B|}_{\rm ribbon}\sim 384\ {\rm G}$,
$S_{\rm ribbon}\sim 5210\ {\rm MSH}$
(assuming the area ratio,
$S_{\rm ribbon}/S_{\rm spot}$, of 85\%),
and $d_{\rm ribbon}$ being, say, 80 Mm,
the estimated magnetic energy
amounts to
$1.5\times 10^{34}\ {\rm erg}$.
Although
what fraction
is converted to the flare energy
is not clear,
the above results indicate
the possibility
that the flare energy
of such gigantic ARs
may be up to of the order
of $10^{34}\ {\rm erg}$.

\begin{figure}
  \begin{center}
    \includegraphics{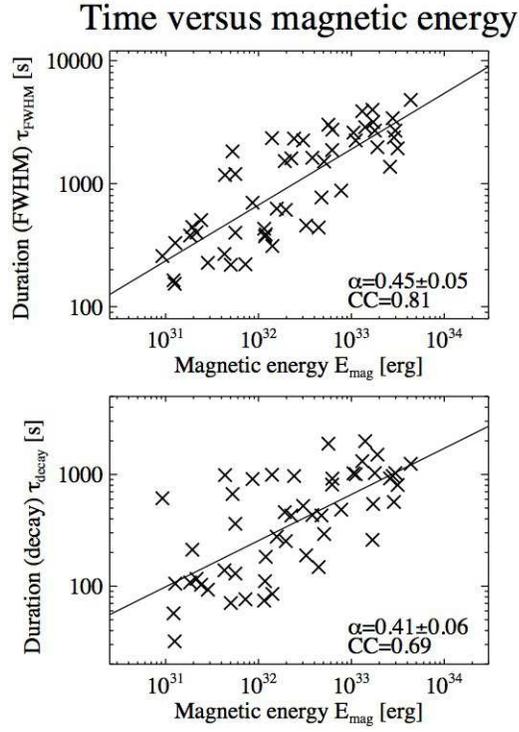}
  \end{center}
  \caption{Scatter plots
    of flare time scales
    $\tau_{\rm FWHM}$ and $\tau_{\rm decay}$
    versus magnetic energy $E_{\rm mag}$
    (equation (\ref{eq:emag})).
    Straight lines show
    the results of linear fitting
    to the log-log plots.
    \label{fig:emag}}
\end{figure}

Figure \ref{fig:emag} compares
the time scales of the flare,
$\tau_{\rm flare}$,
and the magnetic energy
given by equation (\ref{eq:emag}),
$E_{\rm mag}$.
The scatter plots show
the proportionalities of
$\tau_{\rm FWHM}\propto E_{\rm mag}^{0.45\pm0.05}$
(correlation coefficient $CC=0.81$)
and $\tau_{\rm decay}\propto E_{\rm mag}^{0.41\pm0.06}$
($CC=0.69$),
which is surprisingly consistent with
the results of the superflare analysis
by \citet{mae15},
$\tau_{\rm flare}\propto E_{\rm flare}^{0.39\pm0.03}$,
where $\tau_{\rm flare}$ and $E_{\rm flare}$
are the $e$-folding decay time and bolometric energy,
respectively.
Note that their values are
measured from {\it Kepler}'s photometric data
that covers
from 4200 to 9000\ {\AA},
i.e., the optical regime.
\citet{mae15} explained this proportionality
by combining the two relations,
$\tau_{\rm flare}\sim\tau_{\rm A}/M_{\rm A}\sim L/V_{\rm A}/M_{\rm A}\propto L$
(Section \ref{sec:interpretation})
and $E_{\rm flare}\sim fE_{\rm mag}
\sim fB^{2}L^{3}/(8\pi)\propto L^{3}$\footnote{In
their order-of-magnitude estimate,
\citet{mae15} make the ad-hoc assumption that
the sunspot field strength $B$
does not vary much
for different events
and is typically of the same order,
1000 G.}\footnote{Many flare analyses
are based on
the simple assumption
that the flare energy $E_{\rm flare}$
scales with the peak SXR brightness $F_{\rm SXR}$.
However,
the low correlations
between the SXR brightness
and the flare parameters
in Table \ref{tab:power}
(e.g., $CC=0.23$
for $F_{\rm SXR}$ versus $S_{\rm ribbon}$)
may indicate that
the assumption
is not necessarily the case.}
to give $\tau_{\rm flare}\propto E_{\rm flare}^{1/3}$.
However,
from equations (\ref{eq:tflare_dribbon}),
(\ref{eq:tflare_sribbon}), and (\ref{eq:emag}),
one can also derive the relation
$E_{\rm mag}\propto S_{\rm ribbon}d_{\rm ribbon}\propto\tau_{\rm flare}^{2}$,
which may suggest that
the time-energy relation is
$\tau_{\rm flare}\propto E_{\rm mag}^{1/2}$.
Still,
the consistent proportionalities
suggest the existence of the common physical origin
between the solar and stellar flares
\citep{shi99,shi02}.

\section{Conclusion}\label{sec:conclusion}

In this study,
we have examined
all 51 $\ge$M5.0-class,
on-disk ($\le$45$^{\circ}$ from disk center) events,
emanating from 29 ARs,
in the period
of May 2010
to April 2016,
i.e.,
six years
from the activity minimum
of Solar Cycle 24.

Out of the 29 ARs,
24 regions
(83\%)
showed $\delta$-spot configurations,
while
three regions
violated Hale's polarity rule
at the instant
of flare occurrence.
The 51 flare events consist of
20 X- and 31 M-class events.

With the aim to find
the physical parameters
that dictate the GOES light curves,
we systematically surveyed
the correlations
between GOES parameters
(time scales and peak flux)
and AR and flare parameters
(spot size, ribbon size, etc.)
for the 51 events.
The strongest correlations were
obtained for $\tau_{\rm flare}$
(i.e., $\tau_{\rm FWHM}$ and $\tau_{\rm decay}$)
versus $d_{\rm ribbon}$,
$|\Phi|_{\rm ribbon}$, and $S_{\rm ribbon}$,
and all these relations
showed approximately linear correlations.

The first relation,
$\tau_{\rm flare}\propto d_{\rm ribbon}$,
can be explained
by assuming that
(1) the distance between the ribbon composites
in the positive and negative polarities,
$d_{\rm ribbon}$,
represents the length of the reconnected (post-flare) loops,
$L$,
and (2) the flare duration,
$\tau_{\rm flare}$,
is dominated
by the reconnection time,
$\tau_{\rm rec}$,
which should be
related to
the Alfv\'{e}n transit time
over the loop length,
$\tau_{\rm A}\equiv L/V_{\rm A}$.
Then,
we obtain the relation
$\tau_{\rm flare}\sim\tau_{\rm rec}\sim \tau_{\rm A}/M_{\rm A}
\sim L/V_{\rm A}/M_{\rm A}\propto L\propto d_{\rm ribbon}$.
To further investigate this proportionality
with considering the thermal processes,
however,
we may need the help
of loop simulations
including thermodynamics,
because what we observed
is a superposition
of elementary flare loops
and each flare loop undergoes
several stages of thermal processes.

The other two proportionalities,
$\tau_{\rm flare}\propto |\Phi|_{\rm ribbon}$
and $\tau_{\rm flare}\propto S_{\rm ribbon}$,
may be easier to understand.
The former simply shows that
as more magnetic flux is involved,
the reconnection processes continue longer.
The latter may also be accepted
if we assume that the strength of the field lines
are not so different
among the events.

The largest-magnitude,
or longest-duration flares
do not necessarily
produce CMEs.
Although this is obvious
when considering
the perfect example of AR 12192,
the statistical analysis
clearly shows
the general trend that
the non-eruptive events
have smaller $S_{\rm ribbon}/S_{\rm spot}$
and $|\Phi|_{\rm ribbon}/|\Phi|_{\rm AR}$,
which may indicate that
in the non-eruptive regions,
the existence of
embedding field inhibits
CME eruption.
Therefore,
we can speculate that
the relative structural relation
between the flaring region
and the entire AR
is a key to determine
whether the flare becomes eruptive or not.

Most of the
51 flare events under study
originated from the interiors of active regions
Only two events are
not from inside the ARs
but from the boundaries
between separated, independent ARs.
The first group can be subdivided
into three categories,
``spot-spot,'' ``spot-satellite,'' and ``quadrupole.''
The latter,
the ``inter-AR'' group,
shows us that
high-M or
even X-class events can be produced
without strong-field, high-gradient PILs.
The representative event may be
the X1.2 flare
from between ARs 11944 and 11943.
Several scenarios were suggested
in this paper to model
the formation of the above ARs.
These should be examined
through systematic survey
using flux emergence simulations,
which we shall leave for
future research.

The historical record
of a gigantic sunspot group,
RGO 14585,
allows us to know that
even the largest, fragmented ARs
can produce massive flare eruptions
with AR-sized flare ribbons.
The estimation of $d_{\rm ribbon}$
and $S_{\rm ribbon}/S_{\rm spot}$
suggests that
the great flare of RGO 14585
is a long-duration event
with a CME eruption,
which is in line with
the observational facts.
Perhaps
this event points to
the possibility
of the eruption
of even larger ARs.
Estimations suggest that
an AR of $2\times 10^{23}\ {\rm Mx}$
is likely to have existed
and that
if it is flaring,
it could produce superflares
with an energy
of order of $10^{34}\ {\rm erg}$.

Finally,
we found the correlations
of $\tau_{\rm flare}\propto E_{\rm mag}^{0.4}$,
which is well in line with
the stellar flare (superflare) observations.
This clear consistency
favors a
common physical background
for solar and stellar flares.

\acknowledgments

The authors are grateful
to the anonymous referee
for helping us improve the manuscript.
The authors thank ISSI
for the support
of the solar-stellar team.
S.T. would like to thank Dr. Takashi Sakurai
for fruitful discussion
and continuous encouragement.
HMI and AIA are instruments
on board {\it SDO},
a mission for NASA's
Living With a Star program.
This CME catalog is generated and maintained
at the CDAW Data Center by NASA
and The Catholic University of America
in cooperation with the Naval Research Laboratory.
{\it SOHO} is a project of
international cooperation
between ESA and NASA.
The historical Meudon spectroheliograph observations
were digitalized by I. Bual\'{e},
and are available
in the BASS2000 database.
This work was carried out
using the data
from the {\it SDO} HMI/AIA
Joint Science Operations Center
Data Record Management System
and Storage Unit Management System
(JSOC DRMS/SUMS).
This work was partially supported
by JSPS KAKENHI Grant Numbers
JP26887046,
JP16K17671,
and JP15H05814.
K.N. acknowledges support
from EU FP7 Collaborative Project 
``Exploitation of Space Data for
Innovative Helio- and Asteroseismology''
(SPACEINN).


%% for the paper.  Each keyword is check against the master list during
%% copy editing.  Individual instruments can be provided in parentheses,
%% after the keyword, but they are not verified.

\vspace{5mm}
\clearpage

\clearpage

\clearpage

\clearpage

\clearpage

\clearpage

\clearpage

\clearpage

\clearpage

\clearpage

\clearpage

\clearpage

\begin{deluxetable}{cccccCc}
\tabletypesize{\tiny}
\tablecaption{Properties of Flare Events\label{tab:flares1}}
\tablecolumns{7}
\tablewidth{0pt}
\tablehead{
\colhead{Event \#} & \colhead{GOES start} & \colhead{GOES class} &
 \colhead{Position\tablenotemark{a}} &
 \colhead{NOAA \#} &
 \colhead{Classification\tablenotemark{b}} & \colhead{CME}
}
\decimals
\startdata
 1 & SOL2011-02-13T17:28 & M6.6 & S20E05 & 11158 & \beta & N\\
 2 & SOL2011-02-15T01:44 & X2.2 & S20W10 & 11158 & \beta\gamma & Y\\
 3 & SOL2011-03-09T23:13 & X1.5 & N08W11 & 11166 & \beta\gamma\delta & N\\
 4 & SOL2011-07-30T02:04 & M9.3 & N14E35 & 11261 & \beta\gamma\delta & N\\
 5 & SOL2011-08-03T13:17 & M6.0 & N16W30 & 11261 & \beta\gamma\delta & Y\\
 6 & SOL2011-08-04T03:41 & M9.3 & N16W38 & 11261 & \beta\gamma\delta & Y\\
 7 & SOL2011-09-06T01:35 & M5.3 & N13W07 & 11283 & \beta\gamma & Y\\
 8 & SOL2011-09-06T22:12 & X2.1 & N14W18 & 11283 & \beta\gamma & Y\\
 9 & SOL2011-09-07T22:32 & X1.8 & N14W31 & 11283 & \beta\gamma\delta & Y\\
10 & SOL2012-01-23T03:38 & M8.7 & N33W21 & 11402 & \beta\gamma & Y\\
11 & SOL2012-03-07T00:02 & X5.4 & N18E31 & 11429 & \beta\gamma\delta & Y\\
12 & SOL2012-03-07T01:05 & X1.3 & N15E26 & 11429 & \beta\gamma\delta & Y\\
13 & SOL2012-03-09T03:22 & M6.3 & N15W03 & 11429 & \beta\gamma\delta & Y\\
14 & SOL2012-03-10T17:15 & M8.4 & N17W24 & 11429 & \beta\gamma\delta & Y\\
15 & SOL2012-05-10T04:11 & M5.7 & N12E22 & 11476 & \beta\gamma\delta & N\\
16 & SOL2012-07-02T10:43 & M5.6 & S17E06 & 11515 & \beta\gamma & Y\\
17 & SOL2012-07-04T09:47 & M5.3 & S17W18 & 11515 & \beta\gamma\delta & N\\
18 & SOL2012-07-05T11:39 & M6.1 & S18W32 & 11515 & \beta\gamma\delta & N\\
19 & SOL2012-07-12T15:37 & X1.4 & S13W03 & 11520 & \beta\gamma\delta & Y\\
20 & SOL2013-04-11T06:55 & M6.5 & N07E13 & 11719 & \beta\gamma & Y\\
21 & SOL2013-10-24T00:21 & M9.3 & S09E10 & 11877 & \beta\gamma\delta & Y\\
22 & SOL2013-11-01T19:46 & M6.3 & S12E01 & 11884 & \beta\gamma\delta & Y\\
23 & SOL2013-11-03T05:16 & M5.0 & S12W17 & 11884 & \beta\gamma\delta & N\\
24 & SOL2013-11-05T22:07 & X3.3 & S12E44 & 11890 & \beta\gamma\delta & Y\\
25 & SOL2013-11-08T04:20 & X1.1 & S13E13 & 11890 & \beta\gamma\delta & Y\\
26 & SOL2013-11-10T05:08 & X1.1 & S13W13 & 11890 & \beta\gamma\delta & Y\\
27 & SOL2013-12-31T21:45 & M6.4 & S15W36 & 11936 & \beta\gamma\delta & Y\\
28 & SOL2014-01-07T10:07 & M7.2 & S13E13 & 11944 & \beta\gamma\delta & N\\
29 & SOL2014-01-07T18:04 & X1.2 & S12W08 & 11944$^{\ast}$ & \beta\gamma\delta & Y\\
30 & SOL2014-02-04T03:57 & M5.2 & S14W07 & 11967 & \beta\gamma\delta & N\\
31 & SOL2014-03-29T17:35 & X1.0 & N10W32 & 1SOL2017 & \beta\delta & Y\\
32 & SOL2014-04-18T12:31 & M7.3 & S20W34 & 12036 & \beta\gamma & Y\\
33 & SOL2014-09-10T17:21 & X1.6 & N11E05 & 12158 & \beta\gamma\delta & Y\\
34 & SOL2014-09-28T02:39 & M5.1 & S13W23 & 12173$^{\ast}$ & \beta & Y\\
35 & SOL2014-10-22T01:16 & M8.7 & S13E21 & 12192 & \beta\gamma\delta & N\\
36 & SOL2014-10-22T14:02 & X1.6 & S14E13 & 12192 & \beta\gamma\delta & N\\
37 & SOL2014-10-24T21:07 & X3.1 & S22W21 & 12192 & \beta\gamma\delta & N\\
38 & SOL2014-10-25T16:55 & X1.0 & S10W22 & 12192 & \beta\gamma\delta & N\\
39 & SOL2014-10-26T10:04 & X2.0 & S14W37 & 12192 & \beta\gamma\delta & N\\
40 & SOL2014-10-27T00:06 & M7.1 & S12W42 & 12192 & \beta\gamma\delta & N\\
41 & SOL2014-11-07T16:53 & X1.6 & N17E40 & 12205 & \beta\gamma\delta & Y\\
42 & SOL2014-12-04T18:05 & M6.1 & S20W31 & 12222 & \beta\gamma & N\\
43 & SOL2014-12-17T04:25 & M8.7 & S18E08 & 12242 & \beta\gamma\delta & Y\\
44 & SOL2014-12-18T21:41 & M6.9 & S11E10 & 12241 & \beta\gamma\delta & N\\
45 & SOL2014-12-20T00:11 & X1.8 & S19W29 & 12242 & \beta\gamma\delta & Y\\
46 & SOL2015-03-10T03:19 & M5.1 & S15E39 & 12297 & \beta\delta & Y\\
47 & SOL2015-03-11T16:11 & X2.1 & S17E22 & 12297 & \beta\gamma\delta & Y\\
48 & SOL2015-06-22T17:39 & M6.5 & N13W06 & 12371 & \beta\gamma\delta & Y\\
49 & SOL2015-06-25T08:02 & M7.9 & N12W40 & 12371 & \beta\gamma & Y\\
50 & SOL2015-08-24T07:26 & M5.6 & S14E00 & 12403 & \beta\gamma\delta & N\\
51 & SOL2015-09-28T14:53 & M7.6 & S20W28 & 12422 & \beta\gamma\delta & N\\
\enddata
\tablenotetext{a}{Heliographic position.}
\tablenotetext{b}{Mount Wilson sunspot classification
  on the day of the flare occurrence
  provided by NOAA/USAF.
  $\beta$ is assigned to a sunspot group
  which has both positive and negative polarities.
  $\gamma$ indicates that a sunspot group
  has a complex region
  of multiple spots with intermixed polarity.
  $\delta$ indicates that at least one sunspot
  contains opposite polarities
  inside a common penumbra
  separated by no more than 2$^{\circ}$
  in heliographic distance.}
\tablecomments{NOAA number
  with asterisk ($\ast$)
  indicates inter-AR flare.
  Event \#29 occurred
  between NOAA ARs 11944 and 11943,
  and \#34 between ARs 12173 and 12172.}
\end{deluxetable}

\clearpage

\begin{deluxetable}{cRRcRRcRRcRR}
\tabletypesize{\scriptsize}
\tablecaption{Summary of power-law indices
  and correlation coefficients\label{tab:power}}
\tablecolumns{12}
\tablewidth{0pt}
\tablehead{
  \colhead{} &
  \multicolumn{2}{c}{$\tau_{\rm FWHM}$} & \colhead{} &
  \multicolumn{2}{c}{$\tau_{\rm decay}$} & \colhead{} &
  \multicolumn{2}{c}{$F_{\rm SXR}$} & \colhead{} &
  \multicolumn{2}{c}{$V_{\rm CME}$} \\
  \cline{2-3}\cline{5-6}\cline{8-9}\cline{11-12}
  \colhead{} &
  \colhead{$\alpha$} & \colhead{$CC$} & \colhead{} &
  \colhead{$\alpha$} & \colhead{$CC$} & \colhead{} &
  \colhead{$\alpha$} & \colhead{$CC$} & \colhead{} &
  \colhead{$\alpha$} & \colhead{$CC$}
}
\decimals
\startdata
 $S_{\rm spot}$ & 0.43\pm0.17 & 0.35 & & 0.25\pm0.19 & 0.18 & & 0.17\pm0.25 & 0.25 & & 0.34\pm0.17 & 0.34 \\
$|\Phi|_{\rm AR}$ & 0.54\pm0.21 & 0.35 & & -0.65\pm1.74 & -0.05 & & 0.22\pm0.12 & 0.25 & & 0.45\pm0.28 & 0.28\\
$\overline{|B|}_{\rm AR}$ & -0.48\pm1.62 & -0.04 & & 0.30\pm0.23 & 0.18 & & -2.29\pm0.85 & -0.36 & & -2.72\pm1.20 & -0.38 \\
$S_{\rm ribbon}$ & {\bf 1.10\pm0.15} & {\bf 0.72} & & {\bf 1.05\pm0.18} & {\bf 0.64} & & 0.19\pm0.12 & 0.23 & & 0.50\pm0.16 & 0.50 \\
$d_{\rm ribbon}$ & {\bf 0.96\pm0.09} & {\bf 0.83} & & {\bf 0.88\pm0.12} & {\bf 0.71} & & 0.13\pm0.09 & 0.20 & & 0.38\pm0.13 & 0.47 \\
$|\Phi|_{\rm ribbon}$ & {\bf 1.04\pm0.12} & {\bf 0.79} & & {\bf 0.96\pm0.15} & {\bf 0.68} & & 0.28\pm0.10 & 0.37 & & 0.45\pm0.14 & 0.50 \\
$\overline{|B|}_{\rm ribbon}$ & 0.77\pm0.34 & 0.31 & & 0.63\pm0.37 & 0.24 & & 0.48\pm0.19 & 0.34 & & 0.14\pm0.28 & 0.09 \\
\hline
$S_{\rm ribbon}/S_{\rm spot}$ & 0.30\pm0.17 & 0.24 & & 0.45\pm0.18 & 0.34 & & -0.04\pm0.10 & -0.06 & & 0.12\pm0.16 & 0.14 \\
$|\Phi|_{\rm ribbon}/|\Phi|_{\rm AR}$ & 0.79\pm0.17 & 0.54 & & 0.91\pm0.18 & 0.58 & & 0.14\pm0.12 & 0.17 & & 0.36\pm0.17 & 0.37 \\
\hline
$\tau_{\rm FWHM}$ & \nodata & \nodata & & \nodata & \nodata & & \nodata & \nodata & & 0.32\pm0.12 & 0.45 \\
$\tau_{\rm decay}$ & \nodata & \nodata & & \nodata & \nodata & & \nodata & \nodata & & 0.32\pm0.10 & 0.50 \\
$F_{\rm SXR}$ & \nodata & \nodata & & \nodata & \nodata & & \nodata & \nodata & & 0.26\pm0.20 & 0.23 \\
\enddata
\tablecomments{Power-law indices $\alpha$
  and correlation coefficients $CC$
  obtained from various scatter plots are shown.
  Values with higher correlations ($|CC|\ge 0.6$)
  are highlighted with bold face.
  The quantities in the middle part
  ($S_{\rm ribbon}/S_{\rm spot}$
  and $|\Phi|_{\rm ribbon}/|\Phi|_{\rm AR}$)
  are the dimensionless (ratio) parameters,
  which are the combinations
  of the parameters
  in the top part.
  For $V_{\rm CME}$
  the values for the $\tau_{\rm FWHM}$, $\tau_{\rm decay}$,
  and $F_{\rm SXR}$ are also shown
  in the bottom part.
}
\end{deluxetable}

%%%%% Figure and Table %%%%%
\clearpage
\appendix

\section{Target Flares and Measured Parameters}\label{app:list}

Figure \ref{fig:all} lists
the 51 flare events
that we analyzed
in this paper.
Here we show the HMI magnetogram
before the flare onset
as a background
overlaid by the composite flare ribbons.
Ribbon distance is shown 
by a red line
connecting the centroids
of the ribbons
in the positive and negative polarities.
In some events,
we separate the target AR
from neighboring flux concentrations
with thin lines.

Table \ref{tab:flares2} shows
all measured parameters
of the 51 flare events:
GOES parameters
(durations $\tau_{\rm FWHM}$ and $\tau_{\rm decay}$
and GOES flux $F_{\rm SXR}$),
AR parameters
(spot area $S_{\rm spot}$,
total flux $|\Phi|_{\rm AR}$,
and field strength $\overline{|B|}_{\rm AR}$),
flare parameters
(ribbon area $S_{\rm ribbon}$,
distance $d_{\rm ribbon}$,
total flux $|\Phi|_{\rm ribbon}$,
and field strength $\overline{|B|}_{\rm ribbon}$),
and a CME parameter
(CME speed $V_{\rm CME}$).
The maximum, minimum,
median, and standard deviation ($\sigma$)
of each parameter
are summarized
at the bottom of this table.

\clearpage

\begin{figure}
  \begin{center}
    \includegraphics[width=130mm]{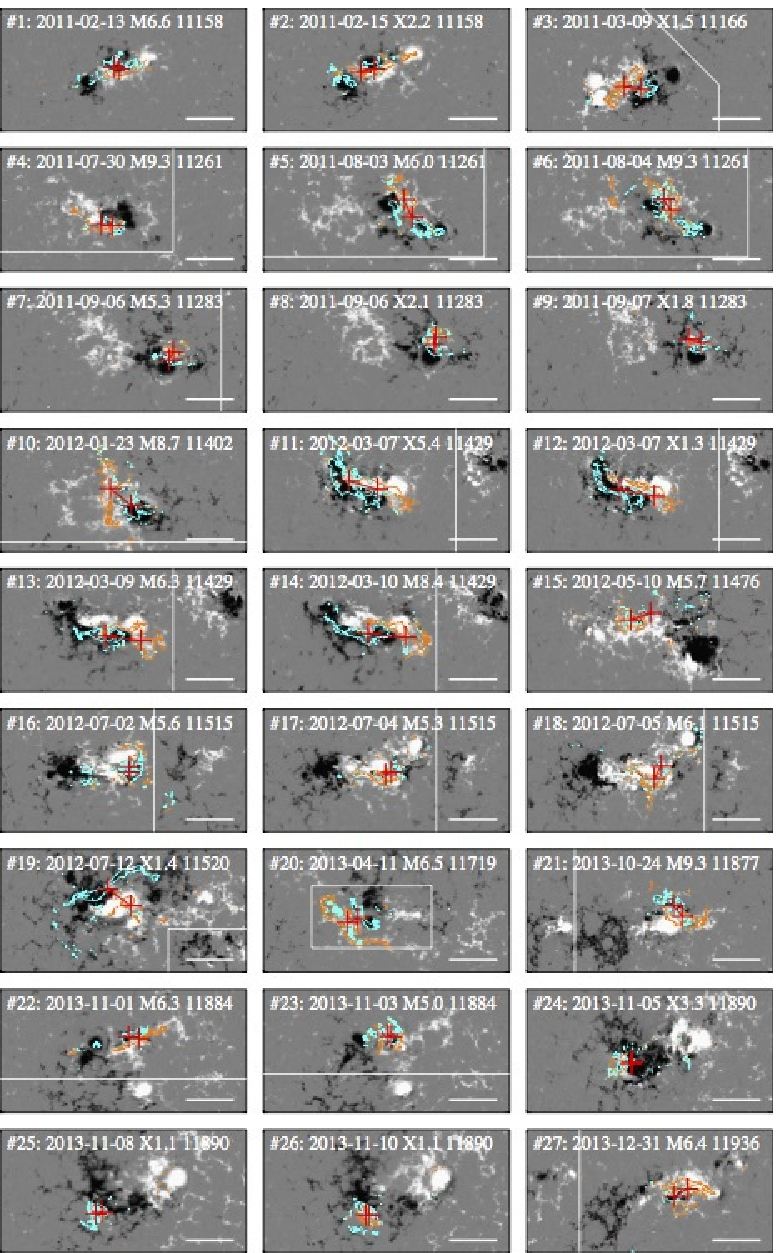}
  \end{center}
  \caption{All flare events
    analyzed in this study.
    Background shows magnetogram
    (saturating at $\pm400\ {\rm G}$),
    over which the composite flare ribbons
    in the positive (orange)
    and negative (turquoise) polarities
    are plotted.
    Centroids of the ribbons
    are denoted by red ``$+$'' signs,
    which are connected
    by a straight line.
    Thick white line
    at the bottom right
    indicates the length of $100\arcsec$,
    while thin white lines separate
    the target AR from neighboring flux concentrations.
    \label{fig:all}}
\end{figure}

\clearpage

\addtocounter{figure}{-1}
\begin{figure}
  \begin{center}
    \includegraphics[width=130mm]{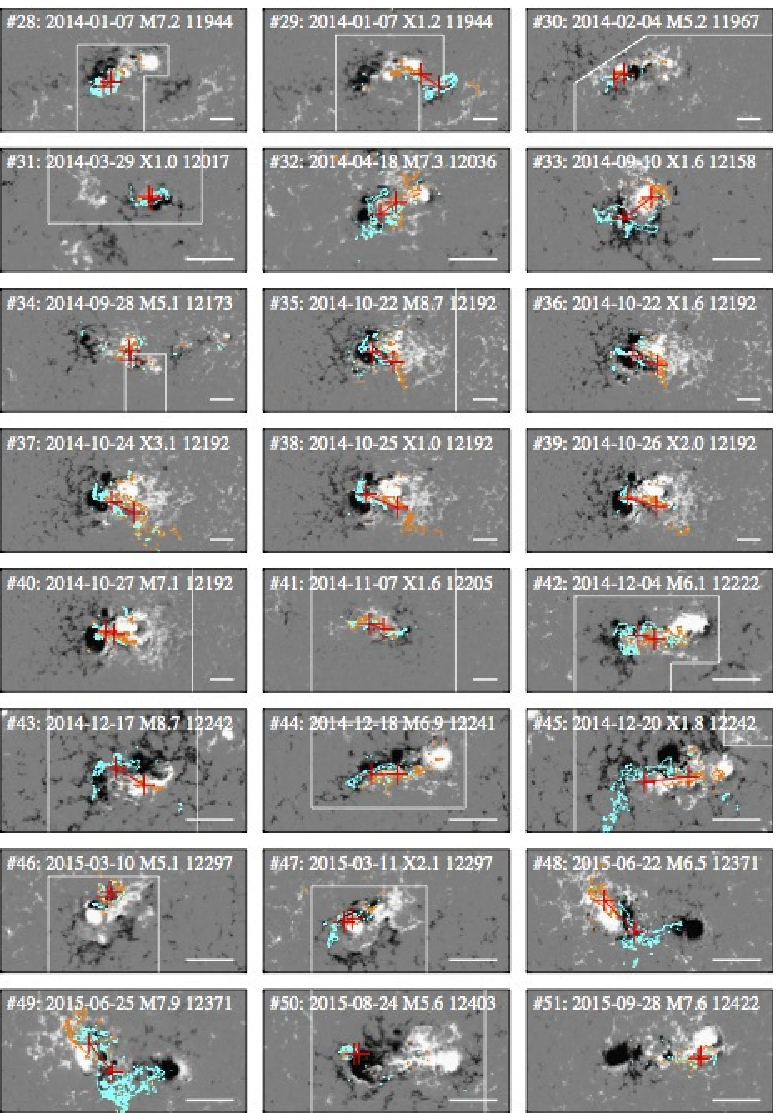}
  \end{center}
  \caption{{\it Continued.}}
\end{figure}

\clearpage

\begin{deluxetable}{cDDDcDDDcDDDDcD}
\tabletypesize{\tiny}
\tablecaption{Measured Parameters of Flare Events\label{tab:flares2}}
\tablecolumns{16}
\tablehead{
\colhead{Event \#} &
 \multicolumn{6}{c}{GOES parameters} & \colhead{} &
 \multicolumn{6}{c}{AR parameters} & \colhead{} &
 \multicolumn{8}{c}{Flare parameters} & \colhead{} &
 \multicolumn{2}{c}{CME} \\
\cline{2-7}\cline{9-14}\cline{16-23}\cline{25-26}
\colhead{} &
 \multicolumn2c{$\tau_{\rm FWHM}$} & \multicolumn2c{$\tau_{\rm decay}$} &
 \multicolumn2c{$F_{\rm SXR}$} & 
 \colhead{} &
 \multicolumn2c{$S_{\rm spot}$} & \multicolumn2c{$|\Phi|_{\rm AR}$} &
 \multicolumn2c{$\overline{|B|}_{\rm AR}$} &
 \colhead{} &
 \multicolumn2c{$S_{\rm ribbon}$} & \multicolumn2c{$d_{\rm ribbon}$} &
 \multicolumn2c{$|\Phi|_{\rm ribbon}$} &
 \multicolumn2c{$\overline{|B|}_{\rm ribbon}$} &
 \colhead{} &
 \multicolumn2c{$V_{\rm CME}$\tablenotemark{a}}\\
\colhead{} &
 \multicolumn2c{(s)} & \multicolumn2c{(s)} &
 \multicolumn2c{($10^{-4}\,{\rm W\,m}^{-2}$)} &
 \colhead{} &
 \multicolumn2c{(MSH)} & \multicolumn2c{($10^{22}\ {\rm Mx}$)} &
 \multicolumn2c{(G)} &
 \colhead{} &
 \multicolumn2c{(MSH)} & \multicolumn2c{(Mm)} &
 \multicolumn2c{($10^{21}\ {\rm Mx}$)} &
 \multicolumn2c{(G)} &
 \colhead{} &
 \multicolumn2c{(${\rm km\ s}^{-1}$)}
}
\decimals
\startdata
 1 &  700. &  910. & 0.66 &&  482. &  2.3 & 770. &&  233. &   9.1 &  2.9 & 411. && \nodata \\
 2 &  772. &  429. & 2.2 &&  678. &  2.8 & 685. &&  380. &  21.1 &  5.7 & 496. &&  669. \\
 3 &  457. &  188. & 1.5 &&  859. &  3.4 & 656. &&  398. &  24.4 &  4.5 & 373. && \nodata \\
 4 &  219. &   71. & 0.93 &&  303. &  2.1 & 662. &&  170. &  16.8 &  1.4 & 270. && \nodata \\
 5 & 2249. &  522. & 0.60 &&  391. &  2.8 & 666. &&  465. &  29.8 &  4.2 & 301. &&  610. \\
 6 &  627. &  277. & 0.93 &&  315. &  2.6 & 634. &&  701. &  18.9 &  4.7 & 222. && 1315. \\
 7 & 1176. &  986. & 0.53 &&  288. &  2.9 & 753. &&  102. &  13.6 &  1.1 & 360. &&  782. \\
 8 &  268. &  139. & 2.1 &&  284. &  2.7 & 668. &&  252. &  10.6 &  2.0 & 259. &&  575. \\
 9 &  399. &  130. & 1.8 &&  220. &  2.5 & 625. &&  126. &  18.7 &  1.2 & 314. &&  792. \\
10 & 2748. &  918. & 0.87 &&  379. &  2.8 & 589. &&  678. &  41.1 &  6.3 & 305. && 2175. \\
11 & 1372. &  922. & 5.4 && 1256. &  3.8 & 588. && 1159. &  44.3 & 16.1 & 459. && 2684. \\
12 & 2699. & 1030. & 1.3 && 1241. &  3.8 & 601. &&  539. &  53.1 &  8.3 & 509. && 1825. \\
13 & 2245. & 1004. & 0.63 &&  960. &  3.8 & 614. &&  661. &  54.0 &  7.2 & 359. &&  950. \\
14 & 3883. & 1315. & 0.84 &&  801. &  3.8 & 601. &&  790. &  56.0 &  8.4 & 350. &&  491. \\
15 &  387. &  183. & 0.57 && 1103. &  5.2 & 730. &&  340. &  31.3 &  2.2 & 216. && \nodata \\
16 &  506. &  103. & 0.56 &&  682. &  3.2 & 703. &&  428. &   8.1 &  2.2 & 170. &&  313. \\
17 &  221. &   76. & 0.53 &&  881. &  4.0 & 695. &&  277. &   9.8 &  2.8 & 333. && \nodata \\
18 &  311. &   85. & 0.61 && 1008. &  4.4 & 652. &&  324. &  20.9 &  2.9 & 294. && \nodata \\
19 & 3985. &  260. & 1.4 && 1369. &  6.8 & 693. &&  562. &  42.5 &  9.2 & 541. &&  885. \\
20 & 1198. &  362. & 0.65 &&  218. &  1.5 & 795. &&  654. &  17.4 &  2.8 & 143. &&  861. \\
21 &  430. &   74. & 0.93 &&  437. &  3.1 & 751. &&  500. &  20.2 &  3.3 & 217. &&  399. \\
22 &  373. &  111. & 0.63 &&  308. &  2.4 & 752. &&  300. &  16.3 &  2.9 & 316. &&  268. \\
23 &  258. &  612. & 0.50 &&  222. &  1.9 & 731. &&  333. &   5.5 &  1.5 & 145. && \nodata \\
24 &  164. &   57. & 3.3 &&  979. &  4.9 & 594. &&  191. &   4.1 &  1.5 & 254. &&  562. \\
25 &  227. &   93. & 1.1 &&  781. &  5.0 & 723. &&  232. &   9.7 &  1.6 & 229. &&  336. \\
26 &  330. &  105. & 1.1 &&  602. &  4.5 & 704. &&  262. &   4.7 &  1.6 & 207. &&  413. \\
27 & 1624. &  433. & 0.64 &&  433. &  3.0 & 668. &&  431. &  23.4 &  5.2 & 395. &&  271. \\
28 & 1526. &  460. & 0.72 && 1679. &  7.6 & 784. &&  745. &  26.9 &  4.5 & 199. && \nodata \\
29 & 2345. &  993. & 1.2 && 1617. &  7.9 & 796. &&  863. &  41.7 &  3.3 & 127. && 1830. \\
30 &  440. &  148. & 0.52 && 1523. &  8.5 & 810. &&  230. &  34.5 &  3.4 & 481. && \nodata \\
31 &  444. &  212. & 1.0 &&  126. &  1.1 & 666. &&  150. &   5.9 &  1.4 & 301. &&  528. \\
32 & 1606. &  428. & 0.73 &&  347. &  2.8 & 674. &&  680. &  29.8 &  4.4 & 215. && 1203. \\
33 & 2875. & 1986. & 1.6 &&  492. &  3.3 & 745. &&  666. &  52.7 &  8.2 & 407. && 1071. \\
34 & 1829. &  668. & 0.51 &&  130. &  1.3 & 790. &&  426. &  32.8 &  1.6 & 125. &&  215. \\
35 & 2697. & 1028. & 0.87 && 2756. & 14.0 & 690. &&  744. &  69.7 & 11.0 & 487. && \nodata \\
36 & 2378. &  567. & 1.6 && 2877. & 15.8 & 655. &&  656. &  79.1 &  9.5 & 479. && \nodata \\
37 & 3404. &  968. & 3.1 && 2781. & 16.6 & 747. && 1639. &  74.4 & 15.3 & 308. && \nodata \\
38 & 4790. & 1244. & 1.0 && 2786. & 16.6 & 744. &&  790. & 105.1 & 11.2 & 466. && \nodata \\
39 & 1937. &  805. & 2.0 && 2572. & 15.8 & 719. &&  645. &  90.8 &  9.2 & 472. && \nodata \\
40 & 1514. &  293. & 0.71 && 2293. & 14.2 & 693. &&  634. &  27.7 &  6.7 & 347. && \nodata \\
41 &  879. &  484. & 1.6 &&  372. &  4.3 & 638. &&  887. &  43.4 &  7.8 & 289. &&  795. \\
42 & 2324. &  968. & 0.61 &&  670. &  2.9 & 661. &&  750. &  25.6 &  5.2 & 229. && \nodata \\
43 & 2601. & 1022. & 0.87 &&  899. &  4.5 & 680. &&  252. &  49.6 &  4.5 & 590. &&  587. \\
44 & 1876. &  812. & 0.69 &&  921. &  3.4 & 705. &&  407. &  39.0 &  5.0 & 402. && \nodata \\
45 & 1976. & 1505. & 1.8 && 1267. &  5.3 & 568. &&  806. &  66.3 &  9.4 & 385. &&  830. \\
46 &  379. &  108. & 0.51 &&  382. &  2.3 & 648. &&  209. &   7.2 &  1.4 & 224. && 1040. \\
47 &  612. &  253. & 2.1 &&  560. &  2.7 & 682. &&  218. &  15.2 &  3.3 & 496. &&  240. \\
48 & 3168. &  544. & 0.65 && 1120. &  4.6 & 699. &&  648. &  64.6 &  8.1 & 412. && 1209. \\
49 & 3017. & 1886. & 0.79 &&  697. &  3.8 & 635. && 1543. &  56.4 &  7.7 & 164. && 1627. \\
50 &  154. &   32. & 0.56 && 1264. &  5.3 & 706. &&  113. &   7.0 &  0.9 & 256. && \nodata \\
51 &  401. &  116. & 0.76 && 1026. &  3.9 & 639. &&  271. &   4.6 &  2.2 & 266. && \nodata \\
\hline
max & 4790. & 1986. & 5.40 && 2877. & 16.6 & 810. && 1639. & 105.1 & 16.1 & 590. && 2684. \\
min &  154. &   32. & 0.50 &&  126. &  1.1 & 568. &&  102. &   4.1 &  0.9 & 125. &&  215. \\
median & 1198. &  433. & 0.87 &&  781. &  3.8 & 685. &&  431. &  26.9 &  4.4 & 308. &&  792. \\
$\sigma$ & 1201. &  487. & 0.89 &&  753. &  4.1 &  60. &&  329. &  24.3 &  3.6 & 119. &&  599. \\
\enddata
\tablenotetext{a}{Non-eruptive events
  are marked with ``$\cdots$''.}
\tablecomments{The maximum, minimum,
  median
  and standard deviation ($\sigma$)
  of each parameter
  are shown at the bottom.
  The values for $V_{\rm CME}$ are
  calculated from the 32 eruptive events.}
\end{deluxetable}

\section{Statistical Tests on the CME-eruptive and Non-eruptive Distributions}\label{app:student}

First,
we compare the spot areas $S_{\rm spot}$
for the CME-eruptive and non-eruptive cases.
Here we use suffix 1 for eruptive distribution
and 2 for non-eruptive group.
From Table \ref{tab:flares2},
the two sample distributions are
\begin{eqnarray}
  S_{\rm spot, 1}&=&[678, 391, 315, 288, 284, 220, 379, 1256, 1241, 960,\nonumber\\
  && 801, 682, 1369, 218, 437, 308, 979, 781, 602, 433,\nonumber\\
  && 1617, 126, 347, 492, 130, 372, 899, 1267, 382, 560,\nonumber\\
  && 1120, 697]\ {\rm MSH}
  \label{eq:b1}
\end{eqnarray}
and
\begin{eqnarray}
  S_{\rm spot, 2}&=&[482, 859, 303, 1103, 881, 1008, 222, 1679, 1523, 2756,\nonumber\\
  && 2877, 2781, 2786, 2572, 2293, 670, 921, 1264, 1026]\ {\rm MSH}
  \label{eq:b2}
\end{eqnarray}
and the numbers of elements
are $n_{1}=32$ and $n_{2}=19$.
Using Student's $t$-test
(Welch's $t$-test),
we examine the null hypothesis
``$\mu_{1}=\mu_{2}$''
and the alternative hypothesis
``$\mu_{1}<\mu_{2}$''
with the one-tailed test,
where $\mu_{1}$ and $\mu_{2}$
are the means of the parent populations.

The statistic $t$ is defined as
\begin{eqnarray}
  t=\frac{\overline{X}_{1} - \overline{X}_{2}}
  {\sqrt{ s_{1}^{2}/n_{1}
      + s_{2}^{2}/n_{2} }},
\end{eqnarray}
where $\overline{X}_{1}$ and $\overline{X}_{2}$
are the means of the two sample distributions,
$s_{1}^{2}=\sum(X_{1i}-\overline{X}_{1})^{2}/(n_{1}-1)$,
and $s_{2}^{2}=\sum(X_{2i}-\overline{X}_{2})^{2}/(n_{2}-1)$.
From (\ref{eq:b1}) and (\ref{eq:b2}),
we obtain $t=-3.732$.

The degrees of freedom $\nu$ is approximated as
\begin{eqnarray}
  \nu=\frac{
      (s_{1}^{2}/n_{1} + s_{2}^{2}/n_{2})^{2}
    }{
      \displaystyle\frac{(s_{1}^{2}/n_{1})^{2}}{n_{1}-1}
      + \displaystyle\frac{(s_{2}^{2}/n_{2})^{2}}{n_{2}-1}
    }
\end{eqnarray}
and is calculated to be $\nu=22.16$,
which is rounded to the nearest integer,
$\nu^{\ast}=22$.

The table relating the test statistics and degrees of freedom
shows that
$t_{0.005}(22)=2.819$
and $t_{0.0005}(22)=3.792$:
\begin{eqnarray}
  -t_{0.0005}(22)<t<-t_{0.005}(22).
\end{eqnarray}
Therefore,
the significant probability
is between 0.0005 and 0.005
and,
at the 99.5\% confidence level,
the null hypothesis is rejected
and we can conclude
that the spot areas of the eruptive events
are smaller than
the non-eruptive events.

Next,
we try the case without AR 12192,
the largest spot group
of the cycle.
The spot areas without the six values for AR 12192 is
\begin{eqnarray}
  S_{\rm spot, 2}&=&[482, 859, 303, 1103, 881, 1008, 222, 1679, 1523, 670,\nonumber\\
  && 921, 1264, 1026]\ {\rm MSH}
  \label{eq:b6}
\end{eqnarray}
and $n_{2}=13$.
Then,
from (\ref{eq:b1}) and (\ref{eq:b6}),
we obtain $t=-1.967$ and $\nu=20.91$ ($\nu^{\ast}=21$).
In this case,
$t_{0.05}(21)=1.721$
and $t_{0.025}(21)=2.080$.
Then, we find that
the probability falls
in between 0.025 and 0.05
and that the null hypothesis
is rejected still
at 95\% confidence.

The AR total flux $|\Phi|_{\rm AR}$
for the eruptive and non-eruptive cases are
\begin{eqnarray}
  |\Phi|_{\rm AR, 1}&=&
  [2.8, 2.8, 2.6, 2.9, 2.7, 2.5, 2.8, 3.8, 3.8, 3.8,\nonumber\\
  && 3.8, 3.2, 6.8, 1.5, 3.1, 2.4, 4.9, 5.0, 4.5, 3.0,\nonumber\\
  && 7.9, 1.1, 2.8, 3.3, 1.3, 4.3, 4.5, 5.3, 2.3, 2.7,\nonumber\\
  && 4.6, 3.8]\times 10^{22}\ {\rm Mx}
\end{eqnarray}
and
\begin{eqnarray}
  |\Phi|_{\rm AR, 2}&=&
  [2.3, 3.4, 2.1, 5.2, 4.0, 4.4, 1.9, 7.6, 8.5, 14.0,\nonumber\\
  && 15.8, 16.6, 16.6, 15.8, 14.2, 2.9, 3.4, 5.3, 3.9]\times 10^{22}\ {\rm Mx}.
\end{eqnarray}
We obtain $t=-3.229$ and $\nu=19.40$ ($\nu^{\ast}=19$).
Since $t_{0.005}(19)=2.861$ and $t_{0.0005}(19)=3.883$,
the probability is between 0.005 and 0.05
and the null hypothesis is rejected
at 99.5\% confidence,
which is the same level
as the test for the spot areas
including AR 12192.


\end{document}